\documentclass[12pt]{article}
\usepackage[a4paper,left=27mm,right=27mm,top=25mm,bottom=27mm]{geometry}
\newif\iffigures
\figurestrue

\newif\iffiguresratepayoff

\newif\ifonlineappendix
\onlineappendixtrue

\newif\ifcomparativestaics

\usepackage[T1]{fontenc}
\usepackage{lmodern}
\usepackage{amsmath,amssymb,amsthm,amsfonts}
\usepackage{enumitem}
\usepackage{hyperref}
\usepackage{mathtools}
\usepackage{microtype}
\usepackage[authoryear]{natbib}
\usepackage{xcolor}
\usepackage{tikz}
\usetikzlibrary{arrows.meta,calc,positioning}
\usepackage{comment}
\usepackage{placeins}
\usepackage{pgfplots}
\usepackage{pgfplotstable}
\usepgfplotslibrary{fillbetween}
\pgfplotsset{compat=1.17}

\hypersetup{
  colorlinks=true,
  linkcolor=black,
  citecolor=black,
  urlcolor=black
}

\numberwithin{equation}{section}

\newtheorem{theorem}{Theorem}[section]
\newtheorem{proposition}{Proposition}[section]
\newtheorem{lemma}{Lemma}[section]

\theoremstyle{definition}
\newtheorem{remark}{Remark}[section]
\newtheorem{definition}{Definition}[section]
\newtheorem{assumption}{Assumption}

\DeclareMathOperator*{\argmax}{arg\,max}

\usepackage{setspace}
\begin{document}
\title{Dynamic Test Design\thanks{This work was funded by European Research Council grant 101165373 (Cetemen). Views and opinions expressed are however those of the authors only and do not necessarily reflect those of the European Union or the European Research Council Executive Agency. Neither the European Union nor the granting authority can be held responsible for them. Dilm\'{e} gratefully acknowledges support from the German Research Foundation (DFG) through CRC TR 224 (Project B02) and under Germany's Excellence Strategy -- EXC 2126/1--390838866 and EXC 2047--390685813. }}
\author{Doruk Cetemen\thanks{LUISS and EIEF;
\protect\href{mailto:dcetemen@luiss.it}{\texttt{dcetemen@luiss.it}}}
\and
Francesc Dilm\'{e}\thanks{University of Bonn; \protect\href{mailto:fdilme@uni-bonn.de}{\texttt{fdilme@uni-bonn.de}}}
}
\maketitle

\begin{quotation}
\renewcommand{\baselinestretch}{1}
\small\selectfont

\renewcommand{\baselinestretch}{1}
\small\selectfont
\noindent 
We study dynamic test design under the threat of manipulation. At each instant, a principal who cannot commit to future actions chooses whether to hire an agent, fire him, or continue costly probation by designing a test. The test provides information about the agent’s persistent type, which is either good or bad. After observing the test choice, the bad type can exert costly effort to manipulate the evidence and mimic the good type. Deterring manipulation imposes an endogenous information budget that limits how intensively the principal can test. We show that, except at beliefs that remain unchanged in the absence of news, equilibrium tests are one-sided: they generate only passes, which trigger hiring, or only failures, which trigger firing. Equilibrium outcomes feature either correct hiring and firing decisions after costly delay or immediate decisions that risk misallocation. With commitment, the principal instead keeps the test constant throughout probation, even when she is nearly certain of the agent’s type.\par

\vspace{2mm}
\noindent \textit{Keywords}: Dynamic testing, Poisson monitoring, optimal stopping 

\vspace{1mm}
\noindent \textit{JEL codes}: C73, D82, D83, D86
\end{quotation}

\newpage

\section{Introduction}

In many long-term relationships, a principal chooses how to learn about an agent's persistent characteristics before taking a consequential action. The agent can, at a cost, shape the evidence generated by this learning process. An employer, for instance, chooses how intensively to evaluate a new hire and whether to search primarily for misconduct or strong performance; the worker chooses how much effort to devote to avoiding adverse signals or building a favorable record before the firm grants permanence or terminates the match. A regulator chooses the frequency and design of audits, while the firm invests in compliance or documentation that shapes the evidence it produces. A digital platform chooses how aggressively to verify contributors, who incur costs to establish reliability or provenance. In each case, the principal chooses the direction and intensity of evidence acquisition, and the agent responds in anticipation of how the resulting evidence will affect future decisions. The principal’s incentive to seek more information and the agent’s incentive to manipulate it therefore depend on the continuation stakes attached to the evidence.


The two sides of the problem have largely been analyzed separately. Models of flexible learning study what a decision maker chooses to learn over time from exogenously available information sources \citep{che2019optimal,zhong2022optimal}. Models of dynamic signaling study how an agent manipulates evidence generated by a technologically fixed signal process \citep{dilme2019dynamic,ekmekci2022learning,cetemen2024dynamic}. 

In this paper we combine the two: the principal chooses the information process and the agent decides how much effort to devote to manipulating the evidence. Our results identify a  trade-off between accuracy and delay. Probation eliminates allocation errors on path, but the direction of monitoring determines who bears the cost of waiting: pass-only monitoring recognizes good types sooner and leaves bad types waiting, whereas fail-only monitoring does the reverse. Thus, monitoring design governs not only what is learned, but also how the delay required for learning is allocated across types.

We study this interaction in a continuous-time principal--agent model in which the agent's persistent type is either good or bad. The good type always exerts high effort. The bad type can shirk at zero cost or imitate the good type at a flow cost. At each posterior belief, the principal either fires, hires, or continues probation, in which case she chooses the arrival rates of two conclusive signals. A pass arrives only under high effort and triggers hiring; a fail arrives only under low effort and triggers firing. The principal incurs a flow cost while the agent is on probation. The principal does not value effort per se: she wants to hire the good type and fire the bad type, while both types want to be hired.

The central friction is that since tests depend on the agents' effort, learning requires the bad type to shirk, yet sufficiently informative monitoring induces him to imitate the good type. A stronger pass test raises the return to imitation, whereas a stronger fail test raises the cost of shirking.  Accordingly, incentive compatibility imposes an endogenous information budget whose weights depend on the bad type's continuation value. As that value evolves, so does the principal's preferred direction of monitoring.

A key observation is that fail-only tests are attractive to the principal when the bad type's continuation value is low, because failure then entails little incremental loss for him. Pass-only tests are instead attractive when his continuation value is high, because passing then yields little incremental gain. This asymmetry yields a sharp implication of strategic manipulation: a fail-only region, in which no news raises the posterior, cannot end in hiring and must instead end at a stationary belief. Symmetrically, a pass-only region cannot end in firing. Thus, conditional on entering probation, monitoring separates types ex post: only the good type is hired and only the bad type is fired, albeit with delay. Throughout either region, the active test becomes less intense over time: the bad type's continuation value rises under fail-only monitoring and falls under pass-only monitoring.


We characterize the unique equilibrium within the class of admissible simple Markov assessments. At sufficiently low and high beliefs, the principal stops immediately, firing at low beliefs and hiring at high beliefs. Fail-only monitoring arises only when manipulation is sufficiently costly: otherwise, deterring imitation requires a fail test too weak to justify probation. Pass-only monitoring arises only when probation is sufficiently cheap, because the principal must be willing to wait for good news. When imitation is costly and probation is cheap, monitoring is fail-only at low beliefs and pass-only at high beliefs, with the two regions joined at an interior stationary belief. For intermediate parameters, probation is one-sided, with either fail-only or pass-only monitoring. When instead imitation is cheap and probation is costly, incentive-compatible learning is not worthwhile, and the principal stops immediately. Remarkably, even though our model features two long-lived players interacting over time, our characterization is in closed form, which eases comparative statics.


We use two benchmarks to isolate the forces shaping the equilibrium. The
first removes manipulation and instead imposes a fixed total testing capacity,
yielding a specialization of \citet{che2019optimal}. Their solution combines
potentially error-prone outward learning near the stopping boundaries
with, when attention is sufficiently cheap, accurate inward learning toward a
stationary belief. Our contribution is to show that the bad type's incentive
constraint eliminates exactly the outward-learning configurations: a pass-only
phase cannot terminate in firing, and a fail-only phase cannot terminate in
hiring. Incentive compatibility therefore replaces these regions with
immediate stopping or inward learning toward a stationary belief. 

The second
benchmark allows the principal to commit ex ante to the entire testing and
stopping policy. Under either nontrivial commitment policy, testing intensities
are constant and no-news beliefs move monotonically toward certainty rather
than toward an interior stationary belief. Commitment matters because it
allows the principal to persist with a policy she would later prefer to
abandon: under maximal fail monitoring, she continues to withhold hiring even
as no news makes the agent arbitrarily likely to be good; under perpetual
pass-only monitoring, she continues to withhold firing even as no news makes
him arbitrarily likely to be bad. Thus, the first benchmark isolates the effect of manipulation, whereas the second isolates the role of commitment
to future testing and stopping.

Two extensions establish the robustness of our results.
First, the baseline equilibria persist when the good type chooses
effort at a lower cost than the bad type: the same tests that deter
imitation also induce high effort from the good type. Second, the
equilibria survive an enlarged menu of monotone imperfect tests,
allowing failures under high effort and passes under low effort.
These errors add effort-independent hiring or firing without relaxing
the incentive constraint on informative testing, so conclusive
testing remains optimal.

Our results identify a general trade-off between accuracy and delay. Probation eliminates allocation errors on path, but the direction of monitoring determines who bears the cost of waiting: pass-only monitoring recognizes good types sooner and leaves bad types waiting, whereas fail-only monitoring does the reverse. Thus, monitoring design governs not only what is learned, but also how the delay required for learning is allocated across types.

\subsection{Related literature}\label{sec:lit}

Empirical work documents both behavioral responses to monitoring and the
importance of continuation stakes. Monitoring changes opportunistic behavior in call centers, public procurement, and teacher attendance \citep{nagin2002monitoring,olken2007monitoring,duflo2012incentives}; among newly hired manufacturing workers, enhanced monitoring raises performance but can also increase turnover \citep{cai2025money}. Behavior also responds to the stakes attached to evaluation. Absence rises after Italian bank employees acquire dismissal protection \citep{ichino2005effect}, whereas greater discretion to dismiss probationary teachers reduces absence \citep{jacob2013effect}. A credible dismissal threat also increases the attrition of low-performing teachers and improves the subsequent performance of those who remain \citep{dee2015incentives}.

Agents also respond to the composition of the evidence used in screening. Accountability incentives induce manipulation of standardized-test outcomes \citep{jacob2003rotten}, while increasing the weight on a performance measure in tenure decisions redirects effort toward that measure and changes who is retained \citep{dinerstein2022screening}. On the principal's side, information acquired after hiring affects separation decisions: employers update in response to objective performance estimates \citep{rockoff2012information}, and realized productivity gradually displaces readily observed prior information \citep{farber1996learning,altonji2001employer,lange2007speed}. Decision makers also redirect attention in response to prior signals and the selection problem they face \citep{bartos2016attention}. Yet these findings do not characterize how a principal dynamically chooses the direction and intensity of monitoring when the agent can reshape the resulting evidence.

Our paper connects flexible information acquisition with dynamic signaling through
endogenous monitoring. Work on flexible learning studies how a decision maker chooses
what to learn before acting. \citet{che2019optimal} analyze the dynamic allocation of
attention across sources biased toward different actions and obtain confirmatory
learning at extreme beliefs and contrarian learning at intermediate beliefs.
\citet{zhong2022optimal} allows a broad class of information processes and shows that
optimal learning can be implemented with Poisson signals. Building on this flexible approach, \citet{georgiadis2024information} studies information acquisition when opportunities to act arrive exogenously, while \citet{ilinov2024information} apply flexible experiment choice to sequential search. These papers explain why optimal
learning may be directional and, in several of these models, why Poisson breakthroughs
arise, but the information flow is chosen by a single decision maker. Here the learner
faces a strategic agent whose hidden action changes the distribution of evidence,
so experiment choice must also satisfy incentive compatibility.

Work on monitoring and test design introduces this incentive problem.
With commitment, \citet{georgiadis2020optimal} and \citet{li2020optimal} endogenize
monitoring in principal--agent problems, while \citet{perezrichet2022test} design
approve--reject tests whose inputs an approval-seeking agent can falsify at a cost. In dynamic contracting, \citet{piskorski2016optimal} study how costly monitoring interacts with performance pay and termination, and \citet{varas2020random} characterize the choice between random inspections and periodic reviews when monitoring has both informational and incentive roles. \citet{bhaskar2025dynamic} combines adverse selection and moral hazard in a
self-enforcing relationship, where monitoring screens low types and shapes
incentives after screening. \citet{dai2022dynamic} and
\citet{wong2023dynamic} instead allow the principal to choose flexible
monitoring technologies that seek evidence tied to rewards or punishments.

As in \citet{dai2022dynamic} and \citet{wong2023dynamic}, our principal
chooses the direction and intensity of monitoring. The key difference is that
our principal cannot commit: our equilibrium notion imposes pointwise
experiment optimality at each nonstationary probation belief rather than
commitment to a future monitoring path. The posterior belief is the state, and the bad type's shirking
constraint creates a belief-dependent information budget. Within the simple
Markov class studied below, this interaction yields one-sided monitoring away
from stationary beliefs and organizes the pass-only and fail-only regions and
the stationary beliefs that connect them.

A parallel literature studies dynamic signaling and reputation under a fixed
monitoring technology. \citet{dilme2019dynamic}, \citet{ekmekci2022learning}, and
\citet{cetemen2024dynamic} study continuous-time signaling or stopping games in which
hidden actions affect a technologically fixed noisy signal. \citet{board2013reputation} and \citet{faingold2011reputation} develop related continuous-time reputation models, while \citet{kuvalekar2020job} study job insecurity with observable behavior and \citet{dilme2025repeated} studies repeated trade when future sellers observe noisy histories of a privately informed buyer’s past transactions. In these models, actions
affect beliefs within an exogenous monitoring technology. We instead endogenize which
kind of conclusive evidence the principal seeks, and at what intensity, subject to the
bad type's incentive constraint. The contribution lies in the interaction between belief-based stopping and this endogenous information budget. 

\paragraph{Roadmap.}
Section~\ref{sec:model} introduces the model, admissible Markov assessments,
posterior dynamics, and continuation values. Section~\ref{sec:incentives}
derives the bad type's incentive constraint, the principal's payoff guarantees,
and her pointwise experiment-choice problem. Section~\ref{sec:equilibrium}
defines equilibrium under interior experiment optimality and characterizes the
phase diagram. Sections~\ref{sec:no-manipulation} and
\ref{sec:commitment} analyze the two benchmarks
. Section~\ref{sec:extensions} discusses strategic effort by the good type
and monotone imperfect tests; Section~\ref{sec:conclusion} concludes.
Appendix~\ref{sec:proofs} contains the baseline proofs and supplementary
results. The Online Appendix
provides additional proofs and extensions.

\section{Model}\label{sec:model}

Time is continuous, \(t\in\mathbb R_+\), and the horizon is infinite. There is a principal (she) and an agent (he). The agent's type is fixed and privately known to the agent:
\[
  \theta\in\{B,G\}, \qquad B<0<G.
\]
The principal's prior belief that the agent is type \(G\) is \(\beta_0\in(0,1)\).

While the agent is on probation, the principal can hire or fire at any moment. If she keeps the agent on probation, she chooses a test
\[
  (\lambda_G,\lambda_B)\in\mathbb R_+^2.
\]
The component \(\lambda_G\) is the arrival rate of a pass under high effort, and \(\lambda_B\) is the arrival rate of a fail under low effort. A pass entails hiring, and a fail entails firing; see Remark \ref{remark_firing_hiring} for a discussion of the role of this assumption.

At each instant $t$ during probation, play proceeds as follows.\vspace{-0.15cm}
\begin{enumerate}[leftmargin=2em]
\item The principal either stops immediately, choosing \(a_t\in\{\mathsf f,\mathsf h\}\), or keeps the agent on probation, choosing \(a_t=\mathsf k\) and a test \((\lambda_G,\lambda_B)\in\mathbb R_+^2\).\vspace{-0.2cm}
\item The agent privately chooses effort \(e_t\in[0,1]\). Effort \(e\) generates a pass at rate \(e\lambda_G\) and a fail at rate \((1-e)\lambda_B\).\vspace{-0.2cm}
\item If a pass occurs, the game ends immediately with hiring. If a fail occurs, the game ends immediately with firing. If neither outcome occurs, the game continues on probation.
\end{enumerate}

The principal's payoff from firing is \(0\). Her payoff from hiring is the agent's type, $\theta$. Keeping the agent on probation costs the principal \(k>0\) per unit of time. The agent receives terminal payoff \(F<0\) if fired and \(H>0\) if hired. While on probation, the agent receives zero flow payoff before effort costs. For the bad type, effort \(e\) costs \(ce\) per unit of time, where \(c>0\). The principal and the agent discount future payoffs at the common rate \(r>0\).

\begin{assumption}\label{ass:good-high-effort}
While on probation, the good type always exerts high effort: \(e_G\equiv1\).
\end{assumption}

Conditional on \(\theta=G\), the behavioral good type passes the test at rate \(\lambda_G\) and fails at rate \(0\). Conditional on \(\theta=B\) and effort \(e\), a pass and a fail arrive at rates \(e\lambda_G\) and \((1-e)\lambda_B\), respectively.\footnote{Equivalently, one can interpret the tests as being informative about type, but the bad type
can manipulate them through costly effort, with $e=1$ replicating the
good type's signal law.}

Assumption~\ref{ass:good-high-effort} keeps the analysis simple by making
the good type behavioral: no good-type effort incentive constraint is imposed,
and no effort cost enters his continuation value. Section~\ref{subsec:strategic-good-type} shows, however, that every equilibrium characterized in Section~\ref{sec:equilibrium} remains valid, with the same policies and thresholds, when the good type is strategic and his effort-cost parameter \(c_H\) satisfies \(0<c_H<c\). The reason is that the bad type's incentive constraint is
the binding restriction on testing: throughout every equilibrium region, the tests
that induce the bad type to shirk satisfy the good type's high-effort incentive
constraint strictly.

\subsection{Simple Markov assessments}\label{ss_simpleMarkov}

We focus on simple Markov assessments that use the public posterior probability \(\beta\) that the agent is type \(G\) as the state variable and prescribe on-path shirking by the bad type, \(e_B\equiv0\). We later justify this restriction within the equilibrium
class studied below.

\begin{definition}
\label{def:markov-assessment}
A \emph{simple Markov assessment} is a triple
\(
\alpha=(a,\lambda_G,\lambda_B)
\)
with the following components.\vspace{-0.15cm}
\begin{enumerate}[leftmargin=1.2em]
    \item \(a:[0,1]\to\{\mathsf f,\mathsf k,\mathsf h\}\) is a Borel measurable action rule. The value \(a(\beta)\) determines whether the principal fires the agent, keeps him on probation, or hires him at belief \(\beta\).\vspace{-0.2cm}

    \item \(\lambda_G,\lambda_B:[0,1]\to\mathbb R_+\) are Borel measurable test rules. At belief \(\beta\), the value \(\lambda_G(\beta)\) is the pass arrival rate under effort \(1\), while \(\lambda_B(\beta)\) is the fail arrival rate under effort \(0\).

\end{enumerate}
\end{definition}

The prescribed test values \(\lambda_G(\beta)\) and \(\lambda_B(\beta)\)
are payoff-irrelevant at beliefs with \(a(\beta)\neq\mathsf k\); they are
retained only to define Borel functions on \([0,1]\). Throughout this
subsection, a simple Markov assessment is interpreted as prescribing on-path shirking by the bad type. The equilibrium conditions below verify shirking
and impose pointwise optimality among tests consistent with shirking at
nonstationary probation beliefs. Thus, at a probation belief \(\beta\),
the good type passes at rate \(\lambda_G(\beta)\), while the bad type fails at
rate \(\lambda_B(\beta)\).

We now impose admissibility conditions on simple Markov assessments to ensure a unique no-news belief path and unique continuation values for the principal and the agent. Similar requirements appear in \cite{klein2011negatively}, \cite{board2013reputation}, and \cite{che2019optimal}.


\begin{definition}
\label{def:admissible-markov-assessment}
Let \(\alpha=(a,\lambda_G,\lambda_B)\) be a simple Markov assessment. Let
\[
\mathcal K:=\{\beta\in[0,1]:a(\beta)=\mathsf k\},
\qquad
\mathcal S:=\{\beta\in\mathcal K:\lambda_G(\beta)=\lambda_B(\beta)\}.
\]
The assessment is admissible if the following conditions hold.\vspace{-0.2cm}
\begin{enumerate}
\item \(\mathcal K\setminus\mathcal S\) is a finite union of open intervals of \(\mathbb R\).\vspace{-0.2cm}
\item The set of stationary probation beliefs \(\mathcal S\) is finite.\vspace{-0.2cm}
\item On each connected component \(I\) of \(\mathcal K\setminus\mathcal S\), the function \(\lambda_B-\lambda_G:I\to\mathbb R\) is locally Lipschitz and has constant nonzero sign, and \(\lambda_G+\lambda_B:I\to\mathbb R_+\) is bounded.\vspace{-0.2cm}

\end{enumerate}
\end{definition}

These regularity requirements define the assessment class studied below. In particular, they exclude unbounded on-path test rates, infinitely many nonstationary probation components or stationary beliefs, and a reversal of the no-news drift within a nonstationary component.

\paragraph{Belief dynamics.} In the probation region of an admissible simple Markov
assessment, belief dynamics follow the usual filtering equation in the absence of news.\footnote{Lemma~\ref{lem:bellman-characterization} in Appendix~\ref{subsec:verification-restrictions} formally establishes that belief dynamics are uniquely determined for admissible assessments, up to the first time the belief exits the nonstationary probation region.} In the absence of a pass or fail, the posterior remains constant at each $\beta\in\mathcal S$. More generally, its drift is
\[
b(\beta)
:=
\beta(1-\beta)(\lambda_B(\beta)-\lambda_G(\beta)).
\]
Thus no news raises the posterior if $\lambda_B(\beta)>\lambda_G(\beta)$, lowers it if $\lambda_G(\beta)>\lambda_B(\beta)$, and leaves it unchanged if $\lambda_B(\beta)=\lambda_G(\beta)$.

\paragraph{Continuation values.} An admissible simple Markov assessment generates a unique pair of continuation values $\Pi:[0,1]\to\mathbb R$ and $V:[0,1]\to\mathbb R$ for the principal and the bad type, respectively.\footnote{See Lemma~\ref{lem:bellman-characterization} in Appendix~\ref{subsec:verification-restrictions}.} At stopping beliefs,
\begin{align*}
&a(\beta)=\mathsf f\ \ \Rightarrow\ \ (\Pi(\beta),V(\beta))=(0,F),\\
&a(\beta)=\mathsf h\ \ \Rightarrow\ \ (\Pi(\beta),V(\beta))=(h(\beta),H),
\end{align*}
where $h(\beta)$ is the principal's expected value from hiring the agent, that is,
\begin{equation}\label{eq:hiring-payoff-beta}
  h(\beta):=\beta G+(1-\beta)B.
\end{equation}
On each nonstationary probation component, the continuation values are locally absolutely continuous and satisfy the Bellman equations almost everywhere:
\begin{align}
r\Pi(\beta)&=-k+b(\beta)\Pi'(\beta)+\beta\lambda_G(\beta)(G-\Pi(\beta))
-(1-\beta)\lambda_B(\beta)\Pi(\beta),
\label{eq:principal-assessment-bellman}\\
rV(\beta)&=b(\beta)V'(\beta)+\lambda_B(\beta)(F-V(\beta)).
\label{eq:bad-assessment-bellman}
\end{align}

%

The next section takes these continuation values as given and asks which tests make shirking optimal for the bad type and which of these are optimal for the principal.

\section{Incentive and optimality conditions}\label{sec:incentives}

This section derives the local incentive and optimality conditions used in the equilibrium definition in Section~\ref{sec:equilibrium}.

\subsection{The bad type's incentive constraint}

Fix an admissible assessment \((a,\lambda_G,\lambda_B)\) and let \(V\) denote the bad type's continuation value. At belief \(\beta\in(0,1)\), suppose the principal offers the test \((\ell_G,\ell_B)\), which may be different from the on-path test \((\lambda_G(\beta),\lambda_B(\beta))\). Assume that the principal believes the bad type will shirk under this test. Then, heuristically, the no-news posterior after a small interval of length \(dt>0\) is
\begin{equation}\label{e_hatbeta}
\widehat\beta_{dt}:=
\frac{\beta e^{-\ell_G dt}}
{\beta e^{-\ell_G dt}+(1-\beta)e^{-\ell_B dt}},
\end{equation}
which is lower than \(\beta\) if \(\ell_G>\ell_B\), higher than \(\beta\) if \(\ell_G<\ell_B\), and equal to \(\beta\) if \(\ell_G=\ell_B\). If the bad type shirks over this interval, his payoff is
\[
\ell_Bdt\,F
+(1-\ell_Bdt)(1-rdt)
V(\widehat\beta_{dt})
+o(dt).
\]
If instead he exerts high effort over the interval, his payoff is
\[
-cdt+
\ell_Gdt\,H
+(1-\ell_Gdt)(1-rdt)
V(\widehat\beta_{dt})
+o(dt).
\]
Shirking is then locally optimal only if the former expression is at least the latter.\footnote{For the induced value \(V\), Lemma~\ref{lem:bellman-characterization} supplies
the required Bellman equation and value matching. If \eqref{eq:bad-ic} holds
for the prescribed test at every probation belief,
Lemma~\ref{lem:global-effort-verification} then shows that shirking is optimal
against all progressively measurable effort deviations, not only one-shot deviations.} Dividing the high-minus-low payoff difference by \(dt\) and taking its upper limit as \(dt\downarrow0\) gives the first-order condition\footnote{\(V(\beta^-)\) and \(V(\beta^+)\) denote the left and right
limits of \(V\) at \(\beta\), respectively. When the corresponding
limit does not exist, \(V(\beta^-)\) denotes the left-hand \(\liminf\)
and \(V(\beta^+)\) denotes the right-hand \(\limsup\).
We set \(V(\beta^0)=V(\beta)\).}
\begin{equation}\label{eq:bad-ic}
\ell_G\bigl[H-V(\beta^{\mathrm{sign}(\ell_B-\ell_G)})\bigr]
-
\ell_B\bigl[F-V(\beta^{\mathrm{sign}(\ell_B-\ell_G)})\bigr]
\le c.
\end{equation}

For a given continuation value function \(V\), define the set of first-order shirking-compatible tests at state \(\beta\) by
\begin{equation}\label{eq:admissible-tests}
\mathcal A(\beta;V):=
\{
(\ell_G,\ell_B)\in\mathbb R_+^2:\text{\eqref{eq:bad-ic} holds}
\}.
\end{equation}
At points where \(V\) is continuous and \(F<V(\beta)<H\), this set is the triangle with extreme points
\[
  \mathcal A^{\text{ext}}(\beta;V):=\bigg\{(0,0),\bigg(\frac{c}{H-V(\beta)},0\bigg),
  \bigg(0,\frac{c}{V(\beta)-F}\bigg)\bigg\}.
\]
The tests for which this first-order constraint binds are
\begin{equation}\label{eq:admissible-tests-indifferent}
\mathcal A_0(\beta;V):=
\{
(\ell_G,\ell_B)\in\mathbb R_+^2:\text{\eqref{eq:bad-ic} holds with equality}
\}.
\end{equation}

The binding incentive sets \(\mathcal A_0(\beta;V)\) are depicted in panel~(a) of Figure~\ref{fig:payoffs-and-A0} for a fixed belief \(\beta\) and several continuation values \(V\). At continuity points with \(F<V<H\), each set is the downward-sloping line segment joining \((c/(H-V),0)\) and \((0,c/(V-F))\), while the corresponding incentive-compatible set \(\mathcal A(\beta;V)\) is the triangle below it. All these segments pass through the stationary test \((\lambda_S,\lambda_S)\) described below. As \(V\) increases, the horizontal intercept rises and the vertical intercept falls, so the boundary becomes flatter. Intuitively, a higher continuation value reduces the bad type's gain from passing, \(H-V\), but increases his gain from avoiding failure, \(V-F\).

\iffigures

\begin{figure}
\centering
\begin{minipage}[t]{0.48\textwidth}
\vspace{0pt}
\centering
\begin{tikzpicture}
\begin{axis}[
    width=0.98\linewidth,
    height=0.72\linewidth,
    axis lines=left,
    xmin=0, xmax=4.25,
    ymin=0, ymax=4.25,
    xlabel={$\ell_G$},
    ylabel={$\ell_B$},
    ylabel style={rotate=-90},
    xtick={0,1,2,3,4},
    ytick={0,1,2,3,4},
    xticklabels={$0$,$\lambda_S$,$2\lambda_S$,$3\lambda_S$,$4\lambda_S$},
    yticklabels={$0$,$\lambda_S$,$2\lambda_S$,$3\lambda_S$,$4\lambda_S$},
    tick label style={font=\scriptsize},
    title={$F<V_1<V_2<V_3<H$},
    title style={font=\small},
    legend style={
        draw=none,
        font=\scriptsize,
        at={(0.98,0.98)},
        anchor=north east,
        row sep=-1pt
    },
    legend cell align=left,
    clip=false,
]

\addplot[black,very thick] coordinates {(0,4) (1.333333,0)};
\addlegendentry{$\mathcal A_0(\beta;V_1)$}

\addplot[black,very thick,dashed] coordinates {(0,2) (2,0)};
\addlegendentry{$\mathcal A_0(\beta;V_2)$}

\addplot[black,very thick,dash dot] coordinates {(0,1.333333) (4,0)};
\addlegendentry{$\mathcal A_0(\beta;V_3)$}

\addplot[only marks,mark=*,mark size=1.9pt] coordinates {(1,1)};
\draw[-{Stealth[length=1.7mm]}] (axis cs:2.12,1.42) -- (axis cs:1.05,1.03);
\node[anchor=west,font=\scriptsize] at (axis cs:2.12,1.42)
    {$(\lambda_S,\lambda_S)$};

\end{axis}
\end{tikzpicture}

\smallskip
{\small\textbf{(a)} Binding incentive sets.}
\end{minipage}
\hfill
\begin{minipage}[t]{0.48\textwidth}
\vspace{0pt}
\centering
\begin{tikzpicture}
\begin{axis}[
    width=0.98\linewidth,
    height=0.72\linewidth,
    xmin=0, xmax=1,
    ymin=-1.2, ymax=1.2,
    axis lines=middle,
    axis line style={->},
    xlabel={$\beta$},
    xlabel style={at={(axis description cs:0.5,-0.14)},anchor=north},
    ylabel={payoff},
    ylabel style={at={(axis description cs:0.02,1.02)},anchor=south},
    title={\phantom{$F<V_1<V_2<V_3<H$}},
    title style={font=\small},
    xtick={0,0.25,0.5,0.8,1},
    xticklabels={$0$,$\beta_{S,0}$,$\beta^*$,$\beta_{S,h}$,$1$},
    xticklabel style={font=\scriptsize},
    ytick=\empty,
    clip=false,
    every axis plot/.append style={line width=1.1pt},
]

\addplot[domain=0:1,samples=2] {2.2*x-1.1};
\node[anchor=south west] at (axis cs:0.91,1.03) {$h$};

\addplot[domain=0:1,samples=2,dashed] {1.2*x-0.3};
\node[anchor=north west] at (axis cs:0.91,0.73) {$\Pi_S$};


\addplot[only marks,mark=*,mark size=1.5pt]
    coordinates {(0.25,0) (0.5,0) (0.8,0.66)};
\draw[densely dotted] (axis cs:0.8,0) -- (axis cs:0.8,0.66);

\end{axis}
\end{tikzpicture}

\smallskip
{\small\textbf{(b)} Payoff crossings.}
\end{minipage}

\caption{Payoff benchmarks and binding incentive sets. Panel (a) shows
$\mathcal A_0(\beta;V_i)$ for fixed $\beta$ and
$F\!<\!V_1\!<\!V_2\!<\!V_3\!<\!H$. Panel (b) shows the
pairwise crossings of $0$, $h$, and $\Pi_S$.}
\label{fig:payoffs-and-A0}
\end{figure}
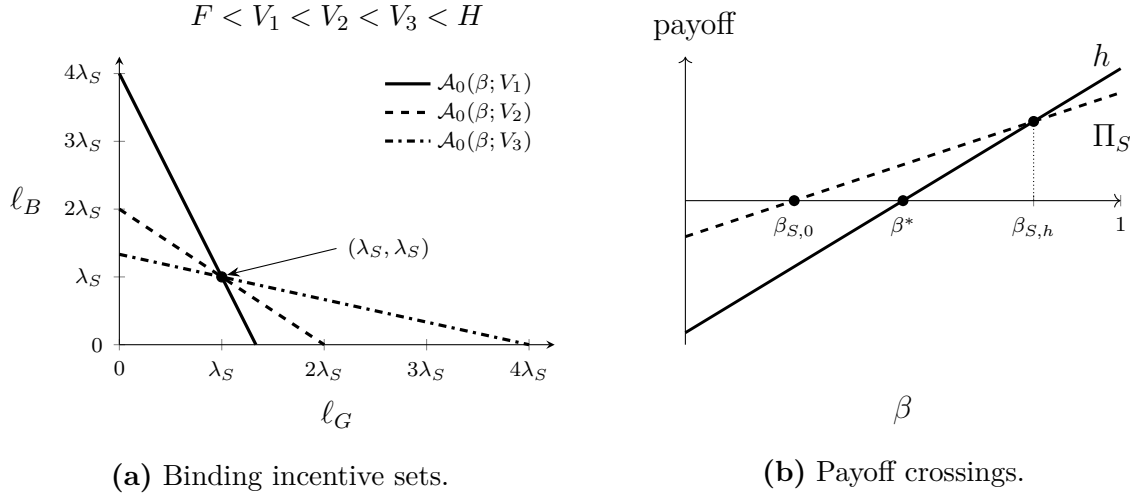

\fi 

Whenever the relevant one-sided limit exists, \eqref{eq:bad-ic} characterizes the bad type's first-order effort incentives under assessed shirking. If \((\ell_G,\ell_B)\notin\mathcal A(\beta;V)\), high effort uniquely maximizes his first-order payoff. If \((\ell_G,\ell_B)\in\mathcal A(\beta;V)\setminus\mathcal A_0(\beta;V)\), zero effort uniquely maximizes it, whereas on \(\mathcal A_0(\beta;V)\) he is indifferent to first order.


\subsection{Minimal guarantees for the principal}\label{ss_minimal_guarantees}

Fix an admissible assessment. Let \(\Pi(\beta)\) denote the principal's continuation value under the assessment. The payoff the principal obtains from firing the agent is \(0\) at each belief \(\beta\). Alternatively, immediate hiring yields \(h(\beta)\), as defined in \eqref{eq:hiring-payoff-beta}.
The static cutoff is
\begin{equation}\label{eq:betastar}
  h(\beta)\geq0
  \qquad\Longleftrightarrow\qquad
  \beta\geq\beta^*=\frac{-B}{G-B}.
\end{equation}

The principal can also keep a stationary test \((\ell,\ell)\) in place until a pass or fail occurs. Stationarity means that the no-news posterior is constant; pass and fail outcomes still arrive and remain informative. Note that
\begin{equation}\label{eq:stationary-rate}
(\ell,\ell)\in\mathcal A(\beta;V)
\qquad\Longleftrightarrow\qquad
\ell\leq\lambda_S:=\frac{c}{H-F}.
\end{equation}
Because the bad type's best response to a stationary test \((\ell,\ell)\) with \(\ell=\lambda_S-\varepsilon\) is shirking for every small \(\varepsilon>0\), the principal can ensure a payoff arbitrarily close to
\begin{equation}\label{eq:stationary-principal-payoff}
  \Pi_S(\beta)
  :=
  \frac{\lambda_S\beta G-k}{r+\lambda_S}.
\end{equation}
Under the on-path selection of shirking at indifference maintained above, the endpoint \(\ell=\lambda_S\) attains \(\Pi_S(\beta)\). We write the bad type's value under this stationary test as
\begin{equation}\label{eq:stationary-agent-payoff}
  V_S:=\frac{\lambda_S F}{r+\lambda_S}.
\end{equation}
The crossings of \(\Pi_S\) with \(0\) and \(h\) are
\begin{equation}\label{eq_def_betaS0Sh}
  \beta_{S,0}
  :=
  \frac{k}{\lambda_SG},
  \qquad
  \beta_{S,h}
  :=
  \frac{-k-(r+\lambda_S)B}
  {r(G-B)-\lambda_SB},
\end{equation}
although \(\beta_{S,0}\) and \(\beta_{S,h}\) need not belong to \([0,1]\). Panel (b) of Figure~\ref{fig:payoffs-and-A0} depicts the three payoff benchmarks.

\subsection{Principal's optimality}

A test that induces high effort by the bad type gives both types the same
signal law and leaves the posterior unchanged. At a continuation value
$\Pi(\beta)\ge\max\{0,h(\beta)\}$, its net first-order payoff gain is
\[
-k-r\Pi(\beta)+\ell_G[h(\beta)-\Pi(\beta)]<0.
\]
Such a response creates no profitable pointwise deviation. The equilibrium
class below imposes pointwise test optimality under assessed shirking.

Fix an admissible assessment and a belief
\(\beta\in\mathcal K\setminus\mathcal S\) at which \(\Pi\) is differentiable. Define the test-dependent part of the principal's first-order payoff gain by
\[
\Psi_P(\beta,\ell_G,\ell_B;\Pi)
:=
\beta(1-\beta)(\ell_B-\ell_G)\Pi'(\beta)
+\beta\ell_G(G-\Pi(\beta))
-(1-\beta)\ell_B\Pi(\beta).
\]
Here and below, pointwise optimality means maximization of the infinitesimal continuation payoff conditional on continuing probation and assessed shirking.

\begin{lemma}\label{lem:extreme-test-reduction}
Let \(\beta\) be an interior point of a probation component of an admissible assessment with continuation values \((\Pi,V)\). Suppose that \(\Pi\) is differentiable at \(\beta\), \(V\) is continuous at \(\beta\), the rates are locally bounded, and $F<V(\beta)<H$. Pointwise optimality among incentive-compatible tests is
equivalent to maximizing \(\Psi_P\) over \(\mathcal A_0(\beta;V)\).
\end{lemma}

Accordingly, at a nonstationary probation belief, pointwise optimality requires
\[
(\lambda_G(\beta),\lambda_B(\beta))
\in
\argmax_{(\ell_G,\ell_B)\in\mathcal A(\beta;V)}
\Psi_P(
\beta,
\ell_G,
\ell_B;
\Pi
).
\]
The continuity requirement imposed below on the principal's value has a direct
economic interpretation. Strictly incentive-compatible one-sided tests let
the principal move the posterior locally in either direction at a cost that
vanishes with the distance traveled.

\subsection{Endpoint restrictions for learning phases}
\label{subsec:endpoint-restrictions-learning-phases}

The principal's payoff guarantees restrict how a nonstationary learning phase can end in finite no-news time. A connected open interval \(I\subseteq\mathcal K\setminus\mathcal S\) is a \emph{pass-learning phase} if \(\lambda_G>\lambda_B\) on \(I\), and a \emph{fail-learning phase} if \(\lambda_B>\lambda_G\) on \(I\). No news moves the posterior toward the lower endpoint of a pass-learning phase and toward the upper endpoint of a fail-learning phase.

\begin{lemma}
\label{lem:learning-phases-wrong-stopping-action}
Consider an admissible assessment with induced continuation values
\((\Pi,V)\) such that
\[
(\lambda_G(\beta),\lambda_B(\beta))\in\mathcal A(\beta;V)
\qquad
\text{for all }\beta\in\mathcal K\setminus\mathcal S,
\]
and
\[
\Pi(\beta)\ge \max\{0,h(\beta),\Pi_S(\beta)\}
\qquad
\text{for all }\beta\in[0,1].
\]
Let \(I=(\beta',\beta'')\) be a connected component of
\(\mathcal K\setminus\mathcal S\). If \(I\) is pass-learning, then its lower
endpoint \(\beta'\), if reached in finite no-news time, cannot be a firing
endpoint. If \(I\) is fail-learning, then its upper endpoint \(\beta''\), if
reached in finite no-news time, cannot be a hiring endpoint.
\end{lemma}

The intuition is simple. Near a firing endpoint, the bad type's continuation
value is close to \(F\), so the binding incentive frontier is steep and permits
a large fail intensity \(\ell_B\). A highly informative fail test is therefore
attractive to the principal, while the bad type's payoff gain from avoiding a failure, \(V(\beta)-F\), is small; hence a pass-learning
phase cannot end in firing. Symmetrically, near a hiring endpoint, the bad
type's continuation value is close to \(H\), so the frontier is flat and
permits a large pass intensity \(\ell_G\). This makes pass information
attractive and rules out a fail-learning phase ending in hiring.

\section{Equilibrium}\label{sec:equilibrium}


\begin{definition}
\label{def:principal-smooth-interior-optimal-mpbe}
An admissible simple Markov assessment is a \emph{simple Markov equilibrium}
if the following conditions hold.

\begin{enumerate}[leftmargin=2em]

\item Every stationary probation belief uses the binding stationary test:
\[
\lambda_G(\beta)=\lambda_B(\beta)=\lambda_S
\qquad
\text{for all }\beta\in\mathcal S.
\]

\item The principal's value is continuous and satisfies the payoff lower bounds:
\[
\Pi\in C([0,1]),
\qquad
\Pi(\beta)\ge
\max\{0,h(\beta),\Pi_S(\beta)\}
\qquad
\text{for all }\beta\in[0,1].
\]

\item At every nonstationary probation belief, \(\Pi\) is differentiable,
the prescribed experiment is pointwise optimal,
\[
(\lambda_G(\beta),\lambda_B(\beta))
\in
\argmax_{(\ell_G,\ell_B)\in\mathcal A(\beta;V)}
\Psi_P(\beta,\ell_G,\ell_B;\Pi),
\qquad
\beta\in\mathcal K\setminus\mathcal S,
\]
and the principal's Bellman equation
\eqref{eq:principal-assessment-bellman} holds pointwise there.

\end{enumerate}

\end{definition}

Throughout, equilibrium means an assessment satisfying
Definition~\ref{def:principal-smooth-interior-optimal-mpbe}. At nonstationary probation beliefs, membership in
the displayed argmax imposes the bad type's incentive constraint; at stationary probation beliefs,
the prescribed test makes this constraint bind.\footnote{Lemma~\ref{lem:global-effort-verification} in the Appendix
therefore implies that the selected on-path shirking behavior is optimal
against every progressively measurable effort deviation.}

In particular, pointwise experiment optimality is imposed only on the
nonstationary probation set. At stationary probation beliefs, the definition prescribes the binding stationary test; the lower-envelope guarantees apply at all beliefs. The assessment specifies on-path shirking, including the selection of zero effort at binding tests, but it does not specify a complete bad-type strategy after
every off-path test or impose optimality against an unrestricted class of
dynamically varying principal controls. The characterization below is confined
to this stated interior-optimality class.

\subsection{Equilibrium characterization}

We now present our main characterization result. In a \emph{fail-only region}, the principal offers tests with $\lambda_B>\lambda_G=0$, and in a \emph{pass-only region}, she offers tests with $\lambda_G>\lambda_B=0$. The thresholds in the statement are obtained in closed form in Section \ref{ss_Regions}.

\begin{theorem}
\label{thm:psio-phase-diagram}
For every parameter vector satisfying the maintained assumptions, there is a unique equilibrium, up to the boundary conventions stated below. There are three thresholds: $k_F$ (independent of $c$), \(c_N(k)\), and \(c_P(k)\). They satisfy \(c_N(k)<c_P(k)\) for every \(k<k_F\) and partition the \((k,c)\)-parameter space as follows:
\begin{enumerate}[leftmargin=2em]

\item[\textnormal{(i)}] \textbf{(No probation)}
If \(c\le c_N(k)\), then \(\mathcal K\setminus\mathcal S\) is empty.

\item[\textnormal{(ii)}] \textbf{(Fail-only)}
If \(k\ge k_F\) and \(c>c_N(k)\), then
\(\mathcal K\setminus\mathcal S\) has exactly one connected component;
this component is a fail-only region, and its right endpoint is a
stationary belief.

\item[\textnormal{(iii)}] \textbf{(Pass-only)}
If \(k<k_F\) and \(c_N(k)<c\le c_P(k)\), then
\(\mathcal K\setminus\mathcal S\) has exactly one connected component;
this component is a pass-only region, and its left endpoint is a
stationary belief.

\item[\textnormal{(iv)}] \textbf{(Fail--pass)}
If \(k<k_F\) and \(c>c_P(k)\), then
\(\mathcal K\setminus\mathcal S\) has exactly two connected components,
one a fail-only region and the other a pass-only region; the right endpoint of
the fail-only region and the left endpoint of the pass-only region coincide at
a stationary belief.

\end{enumerate}

In each case, there is no probation at extreme posteriors: there exists
\(\varepsilon>0\) such that
\(\mathcal K\subset[\varepsilon,1-\varepsilon]\). The equilibrium is unique apart from tests outside the probation set and the principal's stopping choice at
\(\beta^*\) when \(c\le c_N(k)\). At
\(c=c_N(k)\), stationary probation at the isolated belief \(\beta^*\) is also permitted and leaves the principal's value unchanged.
\end{theorem}

Note that the equilibrium tests in $\mathcal K\setminus\mathcal S$ are one-sided, that is, either $\lambda_G=0$ or $\lambda_B=0$. The intuition for one-sidedness is straightforward. The bad type's incentive constraint is a linear testing-capacity constraint, and the principal's Hamiltonian is linear in the pass and fail intensities. Consequently, the principal allocates all available testing capacity to the experiment with the larger return per unit of capacity. A test with both intensities positive can be optimal only when the two experiments tie. Such a tie is equivalent to the principal obtaining the stationary-testing value \(\Pi_S\). This is incompatible with the continuation-value dynamics in a nonstationary region: moving away from a stationary belief raises the bad type's value in a pass-only region and lowers it in a fail-only region, whereas the tie condition can hold only on the opposite side of the stationary belief. Hence the tie cannot occur in \(\mathcal K\setminus\mathcal S\), and the optimal test is necessarily
pass-only or fail-only.

\subsection{Regions}\label{ss_Regions}

\iffigures

\begin{figure}
\centering
\begin{tikzpicture}[scale=1.25,>=stealth]

\coordinate (O)  at (0,0);
\coordinate (P0) at (0,1.2);
\coordinate (X)  at (4,3);
\coordinate (N1) at (7,5.25);
\coordinate (P1) at (7,4.35);

\def\kmax{7}
\def\cmax{5.8}

\fill[gray!12] (O) -- (N1) -- (\kmax,0) -- cycle;
\fill[blue!12] (O) -- (X) -- (P0) -- cycle;
\fill[green!14] (P0) -- (X) -- (4,\cmax) -- (0,\cmax) -- cycle;
\fill[orange!16] (X) -- (N1) -- (\kmax,\cmax) -- (4,\cmax) -- cycle;

\draw[thick] (O) -- (N1);
\draw[thick] node[left] at (3,1.4) {\small $c_N(k)$};
\draw[thick] node[left] at (2,2.3) {\small $c_P(k)$};
\draw[thick] (P0) -- (X);
\draw[thick,dashed] (X) -- (P1);

\draw[dashed] (4,0) -- (4,\cmax);
\node[below] at (4,0) {$k_F$};

\fill (X) circle (1.5pt);
\node[left] at (0,3) {$c^\dagger$};
\draw[dashed] (X) -- (0,3);

\node at (5.6,1.8) {no probation};
\node at (0.69,1.1) {pass-only};
\node at (1.75,4.35) {fail--pass};
\node at (5.25,4.85) {fail-only};

\draw[->] (0,0) -- (\kmax+.35,0) node[right] {$k$};
\draw[->] (0,0) -- (0,\cmax+.35) node[above] {$c$};

\node[left] at (P0) {$-rF$};

\end{tikzpicture}
\caption{Phase diagram in the \((k,c)\)-space. The line \(c_N(k)\) separates the regions with and without nondegenerate probation. For \(k<k_F\), the region between \(c_N(k)\) and \(c_P(k)\) is pass-only, while the region above \(c_P(k)\) is fail--pass. For \(k>k_F\), the region above \(c_N(k)\) is fail-only.}
\label{fig:phase-diagram-k-c}
\end{figure}
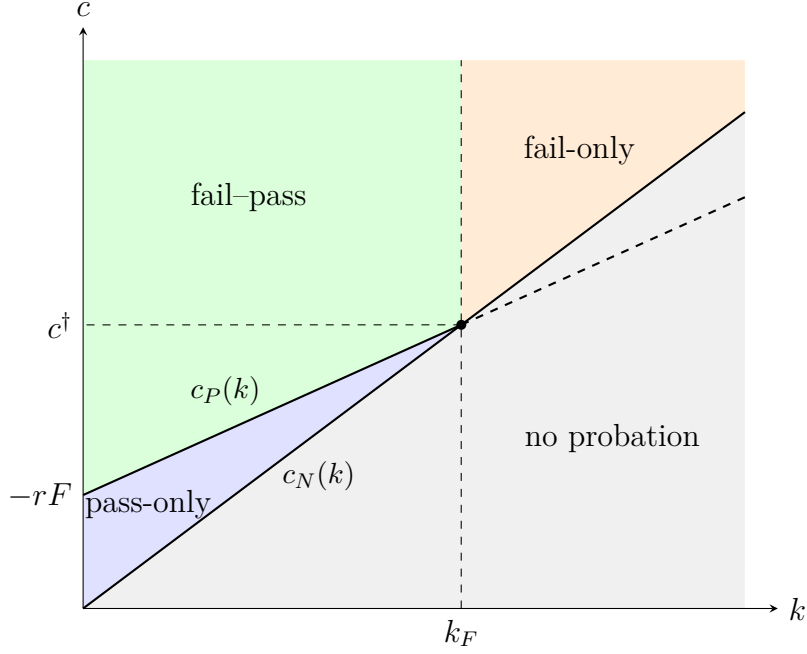

\fi

This section derives the thresholds \(c_N(k)\), \(c_P(k)\), and \(k_F\) that generate the phase diagram in Theorem~\ref{thm:psio-phase-diagram}, and constructs the equilibrium in each resulting region of the \((k,c)\)-parameter space. Figure~\ref{fig:no-news-belief-dynamics} provides a roadmap for these constructions by displaying the stopping regions, probation regions, stationary beliefs, and direction of no-news learning across the four regimes. We consider in turn the no-probation, fail--pass, fail-only, and pass-only cases.

\iffigures

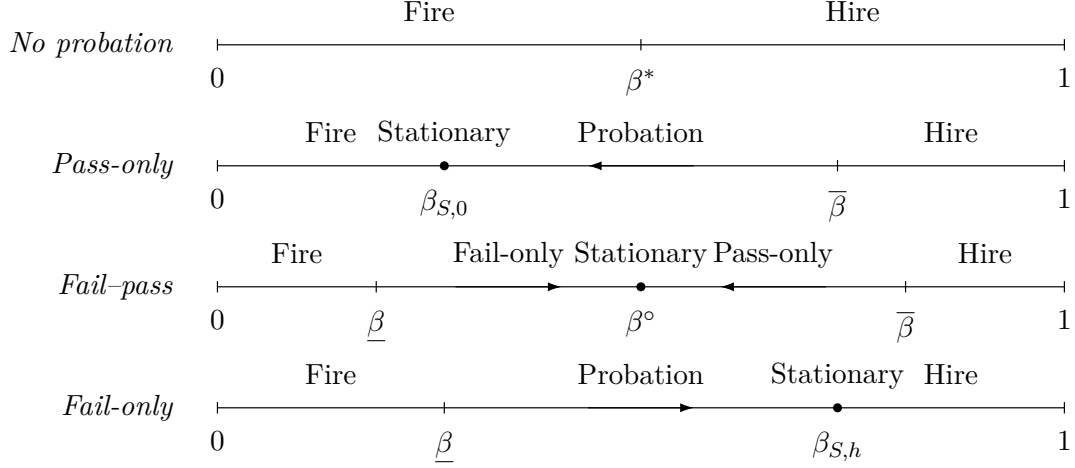
\begin{figure}[t]
\centering
\begin{tikzpicture}[
    x=1cm, y=1cm, font=\small,
    line width=0.45pt,
    >={Latex[length=1.8mm,width=1.2mm]},
    action/.style={above=5pt},
    cutoff/.style={below=5pt},
    stationary/.style={above=5pt, font=\small},
    regime/.style={anchor=east, font=\small\itshape}
]

\foreach \y/\name in {
    4.8/{No probation}, 3.2/{Pass-only},
    1.6/{Fail--pass}, 0/{Fail-only}} {
    \draw (0,\y) -- (11.2,\y);
    \node[regime] at (-0.45,\y) {\name};
    \foreach \x/\lab in {0/0,11.2/1} {
        \draw (\x,{\y-0.075}) -- (\x,{\y+0.075});
        \node[cutoff] at (\x,\y) {$\lab$};
    }
}

\begin{scope}[yshift=4.8cm]
    \draw (5.6,-0.075) -- (5.6,0.075);
    \node[cutoff] at (5.6,0) {$\beta^*$};
    
    \node[action] at (2.8,0) {Fire};
    \node[action] at (8.4,0) {Hire};
\end{scope}

\begin{scope}[yshift=3.2cm]
    \fill (3,0) circle (1.7pt);
    \draw (8.2,-0.075) -- (8.2,0.075);
    \node[cutoff] at (3,0) {$\beta_{S,0}$};
    \node[stationary] at (3,-0.067) {Stationary};
    \node[cutoff] at (8.2,0) {$\overline\beta$};
    \node[action] at (1.5,0) {Fire};
    \node[action] at (5.6,0) {Probation};
    \node[action] at (9.7,0) {Hire};
    \draw[->] (6.3,0) -- (4.9,0);
\end{scope}

\begin{scope}[yshift=1.6cm]
    \foreach \x in {2.1,9.1}
        \draw (\x,-0.075) -- (\x,0.075);
    \fill (5.6,0) circle (1.7pt);
    \node[cutoff] at (2.1,0) {$\underline\beta$};
    \node[cutoff] at (5.6,0) {$\beta^\circ$};
    \node[stationary] at (5.6,-0.067) {Stationary};
    \node[cutoff] at (9.1,0) {$\overline\beta$};
    \node[action] at (1.05,0) {Fire};
    \node[action] at (3.85,-0.067) {Fail-only};
    \node[action] at (7.35,-0.067) {Pass-only};
    \node[action] at (10.15,0) {Hire};
    \draw[->] (3.15,0) -- (4.55,0);
    \draw[->] (8.05,0) -- (6.65,0);
\end{scope}

\draw (3,-0.075) -- (3,0.075);
\fill (8.2,0) circle (1.7pt);
\node[cutoff] at (3,0) {$\underline\beta$};
\node[cutoff] at (8.2,0) {$\beta_{S,h}$};
\node[stationary] at (8.2,-0.067) {Stationary};
\node[action] at (1.5,0) {Fire};
\node[action] at (5.6,0) {Probation};
\node[action] at (9.7,0) {Hire};
\draw[->] (4.9,0) -- (6.3,0);

\end{tikzpicture}
\caption{No-news belief dynamics. Arrows indicate the direction of
belief updating in the absence of a signal; filled dots denote stationary
probation beliefs. Cutoff locations are schematic and are not comparable
across rows. The optional isolated stationary belief at $c=c_N(k)$ is omitted.}
\label{fig:no-news-belief-dynamics}
\end{figure}

\fi 

\paragraph{No probation.}
In case (i) of Theorem~\ref{thm:psio-phase-diagram}, there is no
nondegenerate probation. The threshold \(c_N(k)\) has a direct interpretation:
it is the effort
cost at which the stationary test breaks even at the static cutoff. Indeed, evaluating \(\Pi_S\) at the static cutoff gives
\[
\Pi_S(\beta^*)
=
\frac{G(-B)}{(G-B)[r(H-F)+c]}
[c-c_N(k)],
\]
where
\begin{equation}\label{eq:no-probation-threshold}
  c_N(k)
  :=
\frac{k(H-F)(G-B)}{G(-B)}.
\end{equation}
Thus \(\Pi_S(\beta^*)>0\) if and only if \(c>c_N(k)\). In that case the stationary test gives the principal a strictly positive payoff at the static cutoff, so immediate stopping at the static cutoff cannot satisfy the principal's minimal guarantee.

For \(c<c_N(k)\), the principal fires below the static cutoff and hires above it. 
At \(\beta=\beta^*\), firing and hiring are payoff-equivalent. At the knife edge \(c=c_N(k)\), one may also insert the isolated stationary probation belief
\[
\beta^*=\beta_{S,0}=\beta_{S,h},
\qquad
\lambda_G(\beta^*)=\lambda_B(\beta^*)=\lambda_S,
\]
without changing the principal's value.

\paragraph{Fail--pass regime.}
In case~\textnormal{(iv)}, let \(\beta^\circ\) denote the stationary
posterior. At this belief,
\(\lambda_G(\beta^\circ)=\lambda_B(\beta^\circ)=\lambda_S\),
\(\Pi(\beta^\circ)=\Pi_S(\beta^\circ)\), and
\(V(\beta^\circ)=V_S\), as defined in
\eqref{eq:stationary-rate}, \eqref{eq:stationary-principal-payoff}, and
\eqref{eq:stationary-agent-payoff}. The proof of
Theorem~\ref{thm:psio-phase-diagram} shows that the two extreme one-sided
tests yield the same local return at \(\beta^\circ\):
\[
\underbrace{\beta^\circ\frac{c}{H-V_S}\frac{rG+k}{r+\lambda_S}}_{\text{pass-only local return}}
=
\underbrace{(1-\beta^\circ)\frac{c}{V_S-F}\frac{k}{r+\lambda_S}}_{\text{fail-only local return}}.
\]
Intuitively, at the maximal pass rate
\(c/(H-V_S)\), a pass when the agent is good produces the flow-equivalent
hiring payoff \(rG\) and ends probation, saving the flow cost \(k\). At the
maximal fail rate \(c/(V_S-F)\), a fail when the agent is bad ends probation
and saves \(k\), since the firing payoff is zero. Equating these local returns
gives the stationary posterior
\begin{equation}\label{eq:beta-circ}
\beta^\circ
:=
\frac{k(Hr+c)}{k(Hr+c)-rF(Gr+k)}
\in(0,1).
\end{equation}

For a nondegenerate fail--pass equilibrium to exist, the stationary
posterior $\beta^\circ$ must lie strictly inside the stationary-feasibility interval $(\beta_{S,0},\beta_{S,h})$. Simple algebra shows that
\begin{align}
&\beta^\circ>\beta_{S,0}\quad
\Longleftrightarrow\quad
c>c_P(k):=\frac{k(H-F)}{G}-rF,\label{eq:phase-diagram-thresholds1}\\
&\beta^\circ<\beta_{S,h}
\quad\Longleftrightarrow\quad
k<k_F:=\frac{rBF}{H-F}.
\label{eq:phase-diagram-thresholds2}
\end{align}
Define also $c^\dagger:=c_P(k_F)$. The inverse cost thresholds used below are
\[
k_N(c):=\frac{cG(-B)}{(H-F)(G-B)},
\qquad
k_P(c):=\frac{G(c+rF)}{H-F}.
\]

Because the stationary payoff at \(\beta^\circ\) strictly dominates both
stopping payoffs, the probation set is
\[
\mathcal K=(\underline\beta,\overline\beta),
\qquad
0<\underline\beta<\beta_{S,0}<\beta^\circ
<\beta_{S,h}<\overline\beta<1.
\]
The principal fires below \(\underline\beta\) and hires above \(\overline\beta\). At the stationary belief,
\(
\lambda_G(\beta^\circ)=\lambda_B(\beta^\circ)=\lambda_S
\). On \((\underline\beta,\beta^\circ)\), the system for the fail-only region is solved
to the left from these stationary values.
The lower threshold \(\underline\beta\) is uniquely determined by
\(
\Pi(\underline\beta)=0.
\)
On \((\beta^\circ,\overline\beta)\), the system for the pass-only region is solved
to the right from the same stationary values.
The upper threshold \(\overline\beta\) is uniquely determined by
\(
\Pi(\overline\beta)=h(\overline\beta).
\)


\paragraph{Fail-only one-sided equilibrium.}
In case (ii), there must be a belief
\(\underline\beta<\beta_{S,h}\) such that the principal fires for
\(\beta\leq\underline\beta\), uses fail-only tests for
\(\beta\in(\underline\beta,\beta_{S,h})\), offers the stationary test at
\(\beta_{S,h}\), and hires for \(\beta>\beta_{S,h}\). The lower boundary is uniquely determined by value matching with firing,
\(\Pi(\underline\beta)=0\).



\paragraph{Pass-only one-sided equilibrium.} Case (iii) is analogous to case (ii): there must be a belief
\(\overline\beta>\beta_{S,0}\) such that the principal hires for
\(\beta\geq\overline\beta\), uses pass-only tests for
\(\beta\in(\beta_{S,0},\overline\beta)\), offers the stationary test at
\(\beta_{S,0}\), and fires for \(\beta<\beta_{S,0}\). The upper boundary is uniquely determined by value matching with hiring,
\(\Pi(\overline\beta)=h(\overline\beta)\).


\begin{remark}\label{remark_firing_hiring}
For tractability, we require the principal to hire after a pass and fire after a fail, so that any news arrival ends probation immediately. For fixed continuation values, the bad type's instantaneous best response is $e=0$ when the shirking incentive constraint holds strictly (interior of $\mathcal A(\beta;V)$) and $e=1$ when it is violated (interior of $\mathcal A(\beta;V)^c$). This may not be the case in an alternative formulation where instead the principal retains discretion after observing test outcomes. Consider a pass-only test $(\lambda_G,0)\notin\mathcal A(\beta;V)$, holding the candidate continuation values fixed. Such a test cannot induce $e=1$: a pass would then be uninformative and leave the principal’s continuation decision unchanged, removing the bad type’s incentive to exert costly effort. Nor can it induce $e=0$, since a pass would then be conclusive and lead to hiring, making imitation profitable. 

Our analysis of this alternative formulation, available upon request, shows that the equilibria characterized here remain equilibria for a range of parameter values, but also identifies an open region in which equilibria involve mixed strategies on the equilibrium path. The full analysis therefore requires a richer class of Markov assessments, allowing the principal to randomize over tests and the bad type to exert interior effort on path. Characterizing the resulting mixing probabilities and belief dynamics substantially complicates the analysis while adding limited insight into our central question of how manipulation incentives shape the direction and intensity of testing.

The prescribed consequences of test outcomes can be interpreted as an institutional constraint. For example, a probationary appointment may guarantee confirmation following a successful assessment and require termination following a definitive failure. The employer may retain discretion over the evaluation process and over early stopping while remaining bound by the consequences of a formal assessment once its outcome is observed. The assumption thus captures a limited form of commitment to the consequences of realized test outcomes, without requiring commitment to a future testing policy.
\end{remark}

\section{No-manipulation attention benchmark}
\label{sec:no-manipulation}

To isolate the role of manipulation, suppose that shirking is enforceable and
that the principal allocates a fixed attention capacity between pass and fail
signals:
\begin{equation}
\lambda_G+\lambda_B\leq\bar\lambda,
\qquad \bar\lambda>0.
\label{eq:no-manipulation-capacity}
\end{equation}
This is the attention problem of \citet{che2019optimal}, specialized to our
payoffs. For a clean comparison, choose \(\bar\lambda\) above the largest total
testing intensity used in the strategic equilibrium:
\begin{equation}
\bar\lambda>
\sup_{\beta\in\mathcal K}
\left\{\lambda_G^S(\beta)+\lambda_B^S(\beta)\right\},
\label{eq:attention-capacity-slack}
\end{equation}
where the superscript \(S\) denotes strategic-equilibrium intensities; the
supremum is taken to be zero when \(\mathcal K=\varnothing\). Thus,
imposing the same capacity constraint on the strategic model would leave its
equilibrium unchanged. In \citet{che2019optimal}, the source that produces conclusive evidence for
the good state corresponds to our pass-only monitoring; the other source corresponds to fail-only monitoring.\footnote{To obtain the strictly positive correct-action payoffs required by \cite{che2019optimal}, add \(k/(2r)\) to every terminal payoff and replace the flow cost by \(k/2\). Every policy's discounted value then increases by exactly \(k/(2r)\), including when testing never ends, so optimal testing and stopping are unchanged.}

In this auxiliary benchmark, the principal optimizes jointly over stopping
rules and measurable attention controls adapted to the observed history,
subject to \eqref{eq:no-manipulation-capacity}. The restrictions on
admissible simple Markov assessments in
Definition~\ref{def:admissible-markov-assessment}, used for the strategic
equilibrium analysis, are not imposed on this benchmark. Markov policies
may have isolated switching beliefs; at a repelling switch, their
specification includes a choice of outgoing no-news trajectory.

\begin{proposition}
\label{prop:no-manipulation-attention}
There are thresholds \(0\leq\underline k^A\leq\overline k^A\) such that an
optimal Markov policy has the following form:
\begin{enumerate}[leftmargin=2em]
\item[\textnormal{(i)}] \textbf{(No probation)} If
\(k\geq\overline k^A\), the principal stops immediately.

\item[\textnormal{(ii)}] \textbf{(Pass--fail)} If
\(\underline k^A\leq k<\overline k^A\), probation occurs on
\((\underline\beta^A,\overline\beta^A)\). The principal uses pass-only
monitoring below a switching belief \(\check\beta^A\) and fail-only monitoring
above it. In the absence of news, the posterior moves toward a stopping
boundary. Pass-only and fail-only monitoring use rates
\((\bar\lambda,0)\) and \((0,\bar\lambda)\), respectively.

\item[\textnormal{(iii)}] \textbf{(Pass--fail--pass--fail)} If
\(0<k<\underline k^A\), there are cutoffs
\[
\underline\beta^A<\beta_-^A<\beta_o^A<\beta_+^A<\overline\beta^A
\]
such that monitoring is pass-only on
\((\underline\beta^A,\beta_-^A)\cup(\beta_o^A,\beta_+^A)\), fail-only on
\((\beta_-^A,\beta_o^A)\cup(\beta_+^A,\overline\beta^A)\), and stationary at
\(\beta_o^A=k/(rG+2k)\), where
\(\lambda_G=\lambda_B=\bar\lambda/2\). In the absence of news, the posterior moves toward a stopping boundary in
the two outer regions and toward \(\beta_o^A\) in the two inner regions.
\end{enumerate}
\end{proposition}

\begin{figure}[t]
\centering
\begin{tikzpicture}[
    x=11.2cm, y=1cm, font=\small,
    line width=0.45pt,
    >={Latex[length=1.8mm,width=1.2mm]},
    action/.style={above=6pt},
    cutoff/.style={below=5pt},
    regime/.style={anchor=east, font=\small\itshape},
    short drift/.style={
        ->, >={Latex[length=1mm,width=0.8mm]}
    }
]

\def\slo{0.31}
\def\sswitch{0.34}
\def\shi{0.58}
\def\alo{0.06}
\def\aswitch{0.76}
\def\ahi{0.79}

\foreach \y/\name in {
    1.9/{Strategic}, 0/{No manipulation}} {
    \draw (0,\y) -- (1,\y);
    \node[regime] at (-0.045,\y) {\name};
    \foreach \x in {0,1} {
        \draw (\x,{\y-0.075}) -- (\x,{\y+0.075});
        \node[cutoff] at (\x,\y) {$\x$};
    }
}

\begin{scope}[yshift=1.9cm]
    \foreach \x in {\slo,\shi}
        \draw (\x,-0.075) -- (\x,0.075);

    \node[cutoff,anchor=north east] at (\slo,0)
        {$\underline\beta$};
    \node[cutoff,anchor=north west] at (\sswitch,0)
        {$\beta^\circ$};
    \node[cutoff] at (\shi,0) {$\overline\beta$};

    \node[action] at ({\slo/2-0.02},0) {Fire};
    \node[action,anchor=south east] at (\sswitch,-0.067)
        {Fail-only};
    \node[action] at ({(\sswitch+\shi)/2},-0.067) {Pass-only};
    \node[action] at ({(\shi+1)/2},0) {Hire};

    \draw[short drift] ({\slo+0.003},0)
        -- ({\sswitch-0.008},0);
    \draw[->] (0.52,0) -- (0.43,0);
    \fill (\sswitch,0) circle (1.7pt);
\end{scope}

\foreach \x in {\alo,\ahi}
    \draw (\x,-0.075) -- (\x,0.075);

\node[cutoff] at (\alo,0) {$\underline\beta^A$};
\node[cutoff,anchor=north east] at (\aswitch,0)
    {$\check\beta^A$};
\node[cutoff,anchor=north west] at (\ahi,0)
    {$\overline\beta^A$};

\node[action] at ({\alo/2},0) {Fire};
\node[action] at ({(\alo+\aswitch)/2},0) {Pass-only};
\node[action] at ({(\aswitch+\ahi)/2},0) {Fail-only};
\node[action] at ({(\ahi+1)/2},0) {Hire};

\draw[->] (0.47,0) -- (0.37,0);
\draw[short drift] ({\aswitch+0.008},0)
    -- ({\ahi-0.003},0);
\draw[fill=white] (\aswitch,0) circle (1.9pt);

\end{tikzpicture}
\caption{Strategic and no-manipulation policies on a common belief scale.
Parameters: $G=H=1$, $B=F=-1$, $r=1$, $c=8/5$, $k=1/4$, and
$\bar\lambda=4$. Arrows indicate the direction of no-news belief drift.
The filled point is the strategic stationary belief; the open point is
the no-manipulation benchmark's repelling switch, not a stationary belief.
Cutoff locations are rounded.}
\label{fig:strategic-attention-comparison}
\end{figure}
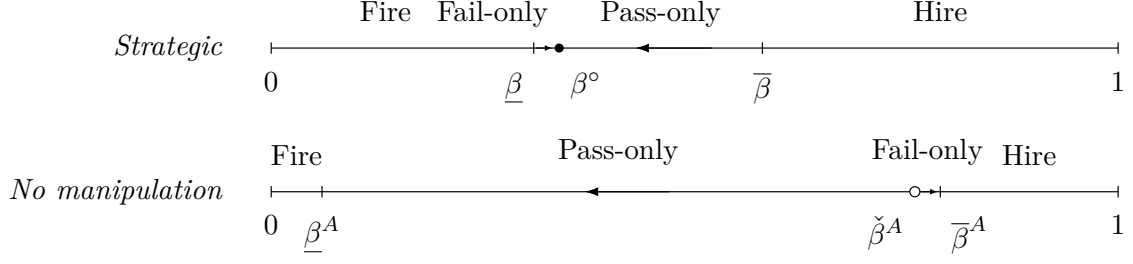
\paragraph{Accuracy.}
Strategic probation classifies both types correctly: no news leads to a
stationary test rather than to a terminal action. The no-manipulation
benchmark is less accurate in its outer regions. Under pass-only monitoring,
a good type is mistakenly fired if no pass arrives before the lower boundary;
under fail-only monitoring, a bad type is mistakenly hired if no fail arrives
before the upper boundary. In the two inner regions of case~\textnormal{(iii)}, the benchmark classifies
both types correctly, as does strategic probation. In
Figure~\ref{fig:strategic-attention-comparison}, strategic monitoring is more
accurate on the common probation interval
\((\underline\beta,\overline\beta)\). The no-manipulation benchmark is more
accurate on
\((\underline\beta^A,\underline\beta)\) and
\((\overline\beta,\overline\beta^A)\), where it continues to learn while the
strategic principal stops; the two models coincide outside
\((\underline\beta^A,\overline\beta^A)\).

\paragraph{Delay.} On the two intervals where the strategic principal stops immediately but
the no-manipulation benchmark continues, the benchmark's additional accuracy
comes with delay. On the common probation interval in
Figure~\ref{fig:strategic-attention-comparison}, a decision-time comparison
must also account for the stopping rules. The no-manipulation benchmark uses
the full capacity \(\bar\lambda\), chosen above the largest strategic total
testing intensity. Its active rate is therefore higher throughout this
interval; wherever both models use pass-only monitoring, this also implies
a higher pass-arrival hazard. Yet, in the absence of news, the benchmark
posterior travels to the lower stopping cutoff \(\underline\beta^A\), whereas
strategic testing reaches \(\beta^\circ\) and then continues until a conclusive
signal arrives. On the small lower part of the common interval, the models
also monitor different signals: strategic monitoring is fail-only, whereas
benchmark monitoring is pass-only. These rate and destination comparisons
alone do not establish whether the models' mean decision times or their
entire decision-time distributions can be ranked.

Within the strategic model, however, the ranking of expected decision times
across types is sharp and depends on the direction of monitoring. Under pass-only monitoring,
the good type may be identified by a pass before beliefs reach the stationary
attachment, whereas the bad type cannot generate a pass and must wait for the
no-news path to reach that attachment and then for a fail. The bad type
therefore waits longer. Under fail-only monitoring the ranking is reversed:
the bad type may be identified by a fail, while the good type must wait for
the stationary test. Hence, in the fail--pass region shown in
Figure~\ref{fig:strategic-attention-comparison}, the good type waits longer
below \(\beta^\circ\) and the bad type waits longer above it. Economically, the
principal's choice of what evidence to seek determines not only the direction
of learning but also which type bears the delay required for perfect
classification.

\section{Commitment benchmark}\label{sec:commitment}

The commitment problem provides a useful benchmark for the equilibrium
analysis: it isolates what the principal can achieve when she chooses the
entire testing and stopping policy ex ante. We formulate the benchmark with the same upper bound \(\bar\lambda\)
on each test intensity. The bound
matters under commitment because the principal may wish to use the most
aggressive feasible test. By contrast, the same bound would not affect the
equilibrium characterization if it exceeded all equilibrium intensities for
the parameter vector under consideration. We assume
\(\bar\lambda\geq c/H\), so the pass-only policy below remains feasible.

\begin{theorem}
\label{thm:optimal-commitment-bounded}
Fix a prior \(\beta\in(0,1)\). Suppose that the principal can commit to a possibly randomized testing and stopping policy, and is subject to the upper bounds \(0\leq\lambda_G\leq\bar\lambda\) and \(0\leq\lambda_B\leq\bar\lambda\)
at every history. Assume that \(\bar\lambda\geq c/H\). At least one of the following policies is optimal:\vspace{-0.2cm}
\begin{enumerate}[label=\textnormal{(\roman*)},leftmargin=2.5em]
\item Immediate firing.\vspace{-0.2cm}
\item Immediate hiring.\vspace{-0.2cm}
\item 
If \(c+rF>0\), maximal fail monitoring with
\(\lambda_B=\bar\lambda\) and
\(\lambda_G=\lambda_G^{\bar\lambda}\), where
\(\lambda_G^{\bar\lambda}\in(0,c/H)\) is the unique constant pass intensity
that makes the bad type indifferent between shirking and exerting high
effort. 
\vspace{-0.2cm}
\item Perpetual pass-only monitoring with
\(\lambda_G=c/H\) and \(\lambda_B=0\), and no stopping in the absence of a pass. 

\end{enumerate}
\end{theorem}

The theorem reduces the commitment problem to a date-zero comparison among
two immediate decisions and the available screening policies. The substantive
new cases are \textnormal{(iii)} and \textnormal{(iv)}. Under policy \textnormal{(iii)}, which is available when \(c+rF>0\), the fail
rate is set at the upper bound and the pass rate is chosen so that the bad
type's incentive constraint binds. Since
\(\lambda_G^{\bar\lambda}<c/H\leq\bar\lambda\), the posterior belief that the
type is good increases toward one as long as no news arrives. Thus the cap binds in the direction that screens the
bad type most aggressively, while the pass intensity keeps him indifferent
between shirking and exerting high effort.

Policy \textnormal{(iv)} has the opposite belief dynamics. With pass-only
monitoring, the posterior belief that the type is good decreases toward zero
as long as no news arrives. A good type is eventually hired at the first pass, whereas a bad type shirks, produces no signal, and remains on probation forever. The principal therefore weighs the cost of keeping a bad type on probation
against the benefit of screening a good type without inducing imitation. 

Note that policies (iii) and (iv) require strong commitment by the principal. Under policy (iii), the absence of news makes her increasingly confident that the agent is good, yet she withholds hiring until a pass occurs. Under policy (iv), the absence of news makes her increasingly confident that the agent is bad, yet she continues probation indefinitely unless a pass occurs. Both policies therefore require her to continue testing even when she would prefer to hire or fire immediately. In practice, such commitment may be difficult to sustain.


\section{Extensions}\label{sec:extensions}

We examine two extensions, each varying one aspect of the baseline
model. The characterized equilibria survive when the good type chooses
effort strategically or, separately, when the test menu includes monotone
imperfect tests. The Online Appendix analyzes the cases of productive
probation and heterogeneous discounting, which preserve the phase-diagram
structure under the stated restrictions, while clarifying how the benefits
and costs of waiting shape monitoring.

\subsection{Strategic effort by the good type}
\label{subsec:strategic-good-type}

The baseline assumes that the good type exerts high effort. This
extension asks whether inducing that effort imposes an additional
restriction on monitoring when both types are strategic. We relax
Assumption~\ref{ass:good-high-effort} by allowing the good type to
choose effort \(e_G\in[0,1]\) at flow cost \(c_He_G\), where
\(0<c_H<c\). All other primitives are unchanged. We retain the
equilibrium notion in
Definition~\ref{def:principal-smooth-interior-optimal-mpbe},
augmented by sequential optimality of the good type's effort.

\begin{proposition}
\label{prop:strategic-good-type}
Every equilibrium characterized in
Theorem~\ref{thm:psio-phase-diagram} admits an extension to the model
with a strategic good type. The principal uses the same stopping and
testing policy, the bad type shirks, and the good type exerts high
effort. The principal's value, the bad type's continuation value,
and all equilibrium thresholds are unchanged. The good type's
high-effort incentive constraint is strict at every probation belief.
\end{proposition}

The result is not immediate from \(c_H<c\). Under the prescribed
policy and effort choices, the good type has a higher continuation
value than the bad type, which reduces his gain from passing even
as it increases his loss from failing. Write these continuation
values as \(V_G\) and \(V\), respectively. The proof in the Online
Appendix shows that, at every
probation belief,
\[
\frac{H-V_G}{c_H}>\frac{H-V}{c},
\qquad
\frac{V_G-F}{c_H}>\frac{V-F}{c}.
\]
Thus, relative to effort cost, both the gain from passing and the
loss from failing are larger for the good type.
Since each baseline test makes the bad type indifferent between
shirking and high effort, it strictly induces high effort by the
good type.

The principal's prescribed tests also remain pointwise optimal.
With pure best responses at off-path tests, the same inequalities
rule out tests that induce high effort only from the bad type.
Tests that induce high effort only from the good type belong to
the baseline incentive-compatible set and generate the same signal
laws as in the baseline model. By contrast, tests that induce the
same effort from both types are uninformative and dominated. The bad type's incentive constraint
therefore remains the binding restriction on information acquisition:
the baseline monitoring policy both deters imitation and provides
the good type with incentives to work.

\subsection{Imperfect tests}
\label{sec:ext-imperfect-tests}

The restriction to conclusive tests raises a design question: could
the principal improve screening by introducing errors that weaken
the bad type's incentive to manipulate? We enlarge the test menu to
allow passes under low effort and failures under high effort,
subject to monotonicity. The baseline equilibria survive because
these errors add random hiring or firing without relaxing the
incentive constraint on the informative part of a test.

Specifically, the principal chooses four rates
\[
(\lambda_G^1,\lambda_G^0,\lambda_B^1,\lambda_B^0)
\in\mathbb R_+^4.
\]
Under effort \(e\in[0,1]\), passes arrive at rate
\(e\lambda_G^1+(1-e)\lambda_G^0\), and failures arrive at rate
\(e\lambda_B^1+(1-e)\lambda_B^0\). We impose
\(\lambda_G^1\ge\lambda_G^0\) and
\(\lambda_B^0\ge\lambda_B^1\): high effort weakly increases the pass
rate and weakly decreases the failure rate. All other baseline
assumptions are unchanged. In particular, the good type is again assumed to exert high effort,
and passes and failures still require hiring and firing, respectively.

\begin{proposition}
\label{prop:monotone-imperfect-tests}
Every equilibrium characterized in
Theorem~\ref{thm:psio-phase-diagram} remains an equilibrium when the
principal's test menu is enlarged to monotone imperfect tests,
within the same equilibrium class and with the same stationary
and boundary conventions. The principal's policy, both continuation
values, and all equilibrium thresholds are unchanged. At every
nonstationary probation belief of these equilibria, tests with
positive error rates are strictly dominated.
\end{proposition}

To see the mechanism, fix the continuation values \((\Pi,V)\) of
a baseline equilibrium and a nonstationary probation belief
\(\beta\). Every monotone test consists of a conclusive test with
rates
\[
\widetilde\lambda_G:=\lambda_G^1-\lambda_G^0\ge0,
\qquad
\widetilde\lambda_B:=\lambda_B^0-\lambda_B^1\ge0,
\]
together with effort-independent pass and failure hazards
\(\lambda_G^0\) and \(\lambda_B^1\). Only the informative components
enter the bad type's shirking constraint:
\[
\widetilde\lambda_G[H-V(\beta)]
+\widetilde\lambda_B[V(\beta)-F]\le c.
\]
Under shirking, the common hazards also cancel from the difference
between the types' total signal rates. Removing them therefore
preserves both local effort incentives and the no-news posterior
drift, holding continuation values fixed.

The common hazards instead replace continuation by hiring or firing
without providing information about type. For a shirking-compatible
test, the principal's first-order payoff effect is
\begin{equation}
\label{eq:imperfect-test-dominance}
\begin{aligned}
\Psi_P^{\mathrm{imp}}
={}&\Psi_P(\beta,\widetilde\lambda_G,
                 \widetilde\lambda_B;\Pi)\\
&+\lambda_G^0[h(\beta)-\Pi(\beta)]
-\lambda_B^1\Pi(\beta).
\end{aligned}
\end{equation}
Since probation strictly dominates both stopping actions throughout
each nonstationary baseline region, either positive error rate
strictly reduces this payoff. Tests that induce high effort from
the bad type give both types the same signal law and remain
uninformative and dominated. The enlarged menu therefore creates
no profitable pointwise testing deviation.

The proof in the Online Appendix also verifies
the stationary and boundary conditions. The result concerns the
principal's choice to introduce errors, not a technology in which
errors are unavoidable. Monotonicity is substantive: it permits
the decomposition into a conclusive test and effort-independent
stopping hazards on which the argument relies.

\section{Conclusion}\label{sec:conclusion}

This paper studies dynamic test design when an agent can manipulate
evidence and the principal cannot commit to future testing or stopping
decisions. The central insight is that information is constrained by the
incentives tests create. More intensive monitoring can induce the bad type
to mimic the good type, generating an endogenous information budget that
depends on his continuation value. The threat of manipulation therefore
shapes monitoring even when no manipulation occurs in equilibrium.

Within the simple Markov equilibrium class studied here, monitoring is
one-sided except at beliefs that remain unchanged in the absence of news.
Fail-only tests are attractive when the bad type's continuation value is
low, because he has less to lose from failure; pass-only tests are
attractive when it is high, because he has less to gain from passing. In either one-sided region, testing becomes less intense
in the absence of news. These results explain why an employer might
concentrate on detecting misconduct rather than recognizing strong
performance, or vice versa. They also show why declining testing intensity
can reflect changing manipulation incentives rather than technological
limitations.

Manipulation also changes the relationship between accuracy and delay.
Without manipulation, the principal may end testing without a conclusive
signal, accepting some risk of misclassification. In the characterized baseline equilibria,
incentive compatibility rules out these stopping patterns: conditional on
entering probation, both types are eventually classified correctly. When
incentive-compatible learning is too costly, the principal instead stops
immediately and risks misallocation. Accurate screening therefore entails
delay, and monitoring determines who bears it: bad types wait longer than
good types under pass-only monitoring, while fail-only monitoring reverses
this ranking. The choice of evidence thus affects both how long suitable
workers wait for confirmation and how long unsuitable workers remain
under evaluation.

The commitment benchmark highlights a separate constraint: a testing
policy that is initially incentive-compatible for the agent can become
unattractive to the principal as information accumulates. Commitment can sustain testing even as beliefs
approach certainty and the principal would otherwise prefer to hire or
fire. This distinction matters for firms whose managers retain discretion
over confirmation and dismissal: an announced evaluation policy need not
bind a manager who prefers to stop rather than prolong costly probation.
Designing a test is therefore distinct from committing to use it until
the prescribed outcome occurs.

The one-sided structure survives heterogeneous discounting and productive
probation under the maintained restrictions. The baseline equilibria also
remain valid when the good type can choose effort strategically and faces
a lower effort cost than the bad type. Allowing monotone imperfect tests likewise leaves these
equilibria intact: introducing failures under high effort or passes under
low effort does not benefit the principal. More broadly, monitoring is
not merely a technology for collecting information. Its design must
account for what evidence can be obtained without inducing manipulation,
how delay is distributed across types, and whether the principal can
credibly maintain testing as her beliefs change.

\appendix

\section{Proofs and supplementary results}
\label{sec:proofs}

\subsection{Value characterization and effort verification}
\label{subsec:verification-restrictions}

This subsection collects two technical results used throughout the analysis.
The first establishes the no-news dynamics and characterizes the induced
continuation values by their Bellman equations. The second turns pointwise
effort optimality into optimality against arbitrary progressively measurable
effort deviations. The proofs for Section~\ref{sec:incentives} follow.

\begin{lemma}
\label{lem:bellman-characterization}
Let $\alpha=(a,\lambda_G,\lambda_B)$ be an admissible simple Markov
assessment prescribing shirking by the bad type. From each nonstationary
belief $\beta$, the no-news path $\hat\beta(\cdot,\beta)$ is uniquely
determined up to its first exit time $\sigma(\beta)$ from the component
containing $\beta$. At a finite exit, write
$\hat\beta(\sigma(\beta),\beta)$ for its limiting endpoint.
The induced continuation values are the unique bounded functions
$\Pi,V:[0,1]\to\mathbb R$ with the following properties.

At stopping beliefs, $(\Pi(\beta),V(\beta))=(h(\beta),H)$ if
$a(\beta)=\mathsf h$, and $(\Pi(\beta),V(\beta))=(0,F)$ if
$a(\beta)=\mathsf f$. At every $\beta\in\mathcal S$,
\[
\Pi(\beta)=\frac{\beta\lambda_G(\beta)G-k}{r+\lambda_G(\beta)},
\qquad
V(\beta)=\frac{\lambda_B(\beta)F}{r+\lambda_B(\beta)}.
\]
On each connected component $I$ of $\mathcal K\setminus\mathcal S$,
both functions are locally absolutely continuous and satisfy
\eqref{eq:principal-assessment-bellman} and
\eqref{eq:bad-assessment-bellman} for almost every $\beta\in I$. Finally, for every
$\beta\in\mathcal K\setminus\mathcal S$ with $\sigma(\beta)<\infty$,
writing $\beta_t:=\hat\beta(t,\beta)$ and $\sigma:=\sigma(\beta)$,
\begin{equation}
\label{eq:assessment-endpoint-value-matching}
\lim_{t\uparrow\sigma}\Pi(\beta_t)=\Pi(\beta_\sigma),
\qquad
\lim_{t\uparrow\sigma}V(\beta_t)=V(\beta_\sigma).
\end{equation}
\end{lemma}

\begin{proof}
Fix $\beta_0\in\mathcal K\setminus\mathcal S$, and let $I$ be its
connected component. Admissibility makes $\lambda_B-\lambda_G$, and hence
$b$, locally Lipschitz on $I$. Picard--Lindel\"of gives a unique maximal
classical path $\hat\beta(\cdot,\beta_0):[0,\sigma(\beta_0))\to I$
satisfying
\begin{equation}
\label{eq:no-news-belief-ode}
\hat\beta(0,\beta_0)=\beta_0,
\qquad
\dot{\hat\beta}(t,\beta_0)=b\bigl(\hat\beta(t,\beta_0)\bigr).
\end{equation}
Since $b$ has constant nonzero sign on $I$, the path is strictly monotone.
Dividing by $\hat\beta(1-\hat\beta)$ gives
\begin{equation}
\label{eq:no-news-log-odds}
\frac{d}{dt}\log\frac{\hat\beta(t,\beta_0)}{1-\hat\beta(t,\beta_0)}
=\lambda_B\bigl(\hat\beta(t,\beta_0)\bigr)
-\lambda_G\bigl(\hat\beta(t,\beta_0)\bigr).
\end{equation}
At a finite exit, monotonicity gives a limit in $\overline I$. An interior
limit would allow extension, contrary to maximality, so the limit lies in
$\partial I$. There the assessment either stops or, by the stationary-belief
convention, keeps the posterior fixed. This absorption is specified separately
from the local Lipschitz condition on the open component. On each compact interval $J\subseteq I$,
continuity and the nonzero sign of $b$ give
$0<m_J\le |b(\beta)|\le M_J<\infty$. Integrating these bounds shows that
the no-news flow and its inverse are Lipschitz while the path stays in $J$.

Immediate stopping gives the stated terminal values. At a stationary belief,
$\lambda_G=\lambda_B$ and the posterior remains fixed; conditioning on the
first signal gives the stated stationary values. Terminal payoffs are
bounded, and discounted probation costs are at most $k/r$, so both induced
continuation values are bounded.

Fix $\beta\in\mathcal K\setminus\mathcal S$, let $I$ be its connected
component, and write $\beta_t:=\hat\beta(t,\beta)$ and
$\sigma:=\sigma(\beta)$. For $0\le t<\sigma$, define the discounted
survival probabilities
\[
\begin{aligned}
D_P(t,\beta)
&:=\exp\left\{-rt-\int_0^t
\bigl[\beta_s\lambda_G(\beta_s)
+(1-\beta_s)\lambda_B(\beta_s)\bigr]ds\right\},\\
D_B(t,\beta)
&:=\exp\left\{-rt-\int_0^t\lambda_B(\beta_s)\,ds\right\}.
\end{aligned}
\]
Under assessed shirking, public pass and fail intensities are
$\beta_t\lambda_G(\beta_t)$ and $(1-\beta_t)\lambda_B(\beta_t)$;
conditional on the bad type, only failure occurs, at rate
$\lambda_B(\beta_t)$. Conditioning on the first signal before any
$0\le T<\sigma$ therefore gives
\begin{align}
\Pi(\beta)
&=\int_0^T D_P(t,\beta)
\bigl[-k+\beta_t\lambda_G(\beta_t)G\bigr]dt
+D_P(T,\beta)\Pi(\beta_T),
\label{eq:principal-pathwise-representation}\\
V(\beta)
&=\int_0^T D_B(t,\beta)\lambda_B(\beta_t)F\,dt
+D_B(T,\beta)V(\beta_T).
\label{eq:bad-pathwise-representation}
\end{align}

On compact subintervals of $I$, these representations and the positive,
absolutely continuous survival factors imply absolute continuity of
$\Pi(\beta_t)$ and $V(\beta_t)$ in time. The local Lipschitz property
of the no-news flow and its inverse then makes both values locally absolutely
continuous in belief, and differentiation yields
\eqref{eq:principal-assessment-bellman}--\eqref{eq:bad-assessment-bellman}
almost everywhere. If $\sigma<\infty$, conditioning up to the exit gives
\eqref{eq:assessment-endpoint-value-matching}: bounded rates imply that,
as the remaining travel time vanishes, the accumulated payoff terms tend
to zero and the discounted survival probabilities tend to one.

Conversely, let $(\widetilde\Pi,\widetilde V)$ be any bounded pair
satisfying the stated conditions. Along each no-news path, multiply its
Bellman equations by $D_P$ and $D_B$, respectively, and integrate. The
chain rule gives
\eqref{eq:principal-pathwise-representation}--\eqref{eq:bad-pathwise-representation}
with the candidate values in place of $(\Pi,V)$. At a finite exit, value
matching and the stopping or stationary formulas determine the terminal
terms, since the reached endpoint lies outside
$\mathcal K\setminus\mathcal S$. At an infinite exit, boundedness and
$D_P(T,\beta),D_B(T,\beta)\le e^{-rT}$ make those terms vanish as
$T\to\infty$. Thus the candidate pair equals the induced values at every
belief, proving uniqueness.
\end{proof}

\begin{lemma}
\label{lem:global-effort-verification}
Let \(\alpha=(a,\lambda_G,\lambda_B)\) be an admissible simple Markov
assessment, and let
\(\bar c>0\) be an effort-cost parameter. Let \(W\) be a bounded candidate
continuation value for an agent whose effort \(e\in[0,1]\) generates a pass at
rate \(e\lambda_G\), a fail at rate \((1-e)\lambda_B\), and flow cost
\(\bar c e\). Suppose that:
\begin{enumerate}[label=\textnormal{(\roman*)},leftmargin=2.5em]
\item \(W=H\) after hiring or a pass and \(W=F\) after firing or a fail;

\item \(W\) is locally absolutely continuous on each nonstationary component
and satisfies value matching at every finite no-news attachment; and

\item at almost every nonstationary probation belief, and algebraically at
every stationary probation belief,
\begin{equation}
\label{eq:controlled-agent-bellman}
rW
=
bW'+\lambda_B(F-W)
+
\max_{u\in[0,1]}u\,\Gamma_W,
\qquad
\Gamma_W
:=
\lambda_G(H-W)-\lambda_B(F-W)-\bar c,
\end{equation}
where the term \(bW'\) is omitted at a stationary belief.
\end{enumerate}
Then every measurable pointwise selector
$
e^*(\beta)\in\argmax_{u\in[0,1]}u\,\Gamma_W(\beta)
$
is optimal against all progressively measurable effort deviations. More
precisely, for every initial probation belief \(\beta_0\) and every
progressively measurable process \(e_t\in[0,1]\),
\begin{equation}
\label{eq:effort-verification-gap}
J_{\bar c}(e;\beta_0)-W(\beta_0)
=
\mathbb E\int_0^{\tau_{\mathrm{end}}}e^{-rs}
\left[
 e_s\Gamma_W(\beta_s)
-
\max_{u\in[0,1]}u\Gamma_W(\beta_s)
\right]ds
\le0,
\end{equation}
where \(J_{\bar c}(e;\beta_0)\) is the agent's expected discounted terminal
payoff net of effort costs and \(\tau_{\mathrm{end}}\) is the first signal or
the assessed stopping time.

In particular, if \(W=V\), \(\bar c=c\), and
\begin{equation}
\label{eq:on-path-global-shirking-condition}
\lambda_G(\beta)\bigl[H-V(\beta)\bigr]
-
\lambda_B(\beta)\bigl[F-V(\beta)\bigr]
\le c
\end{equation}
throughout probation, then \(e_B\equiv0\) is globally optimal. If the
inequality is strict along every state occupied with positive discounted time,
shirking is the unique optimum. Symmetrically, if \(\Gamma_W>0\) throughout
probation, high effort is the unique optimum.
\end{lemma}

\begin{proof}
Fix a progressively measurable effort process \(e_t\in[0,1]\).
Admissibility bounds both intensities on the probation set: there are only
finitely many nonstationary components and stationary beliefs, and the rates
are bounded on each. Under the deviation, the pass and fail intensities are
\(e_t\lambda_G(\beta_t)\) and \((1-e_t)\lambda_B(\beta_t)\), while the
public no-news drift remains \(b\).

Apply the chain rule and the compensator formula to discounted \(W\), less
accumulated effort costs, up to \(\tau_{\mathrm{end}}\wedge n\), for an
integer \(n\ge1\). The controlled Bellman equation makes its predictable
drift
$
e^{-rt}\left[e_t\Gamma_W(\beta_t)
-\max_{u\in[0,1]}u\Gamma_W(\beta_t)\right].
$
The same identity holds at stationary beliefs by the algebraic Bellman
equation. Matching at each reached endpoint cancels the intermediate boundary
terms. Bounded rates and bounded \(W\) make the stopped compensated terms
integrable, so taking expectations gives the stopped version of
\eqref{eq:effort-verification-gap}.

Let \(n\to\infty\). The terminal continuation term vanishes on
\(\{\tau_{\mathrm{end}}=\infty\}\), and the discounted effort cost is
bounded by \(\bar c/r\). The drift integrand is also bounded by a constant
times \(e^{-rt}\). Dominated convergence therefore gives
\eqref{eq:effort-verification-gap}. Its integrand is nonpositive for every
effort process and is zero for a maximizing selector. This proves optimality
and both sign-based conclusions; strict optimality is understood up to
changes on sets of zero time--probability measure before termination.
\end{proof}

\subsection{Proofs of results in Section \ref{sec:incentives}}

\begin{proof}[Proof of Lemma~\ref{lem:extreme-test-reduction}]
For an incentive-compatible test \((\ell_G,\ell_B)\), the principal's
payoff over \(dt\), followed by continuation, is
\[
-kdt+\beta\ell_GGdt+
\bigl[1-\beta\ell_Gdt-(1-\beta)\ell_Bdt\bigr](1-rdt)
\Pi(\widehat\beta_{dt})+o(dt).
\]
Using \(\widehat\beta_{dt}-\beta
=\beta(1-\beta)(\ell_B-\ell_G)dt+o(dt)\), its first-order term is
\(-k-r\Pi+\Psi_P(\beta,\ell_G,\ell_B;\Pi)\). This proves the
Hamiltonian characterization.

Moreover, \(\Pi(\beta)<\beta G\): the principal's discounted terminal
payoff is at most \(G\) conditional on the good type and at most zero
conditional on the bad type, and a strict continuation belief entails a
strictly positive expected probation cost. For every
\(0<\ell\le c/(H-F)\), the common-rate test is feasible and has Hamiltonian
\(\ell(\beta G-\Pi)>0\). The zero test therefore cannot maximize.

Because \(F<V<H\), the feasible set is the triangle with vertices
\((0,0)\), \((c/(H-V),0)\), and \((0,c/(V-F))\). Any nonzero test with slack incentives and a positive Hamiltonian can be
scaled up to increase that Hamiltonian. Thus every maximizer lies on the
edge joining the two nonzero vertices. This is exactly \(\mathcal A_0(\beta;V)\),
and its endpoints are the stated one-sided tests.
\end{proof}

\begin{proof}[Proof of Lemma~\ref{lem:learning-phases-wrong-stopping-action}]
Use posterior log-odds and their inverse,
\begin{equation}\label{eq:log-odds}
x:=\log\frac{\beta}{1-\beta},
\qquad \beta(x):=\frac{e^x}{1+e^x}.
\end{equation}
Bounded rates imply that a finite no-news endpoint lies in \((0,1)\).
For a belief in a component with such an endpoint, let \(\sigma(\beta)\)
be the remaining no-news travel time. This time tends to zero as the endpoint
is approached from within the component.

Suppose first that a pass-learning component ends in firing at its lower
endpoint \(\beta'\). Endpoint matching and the lower envelope imply
\(\Pi\to0\), \(V\to F\), and \(\beta'\le\beta_{S,0}\). Incentive
compatibility gives
\[
\lambda_G\le\lambda_S+
\frac{(\lambda_G-\lambda_B)(V-F)}{H-F},
\qquad \limsup_{\beta\downarrow\beta'}\lambda_G\le\lambda_S.
\]
If \(\beta'<\beta_{S,0}\), the principal's payoff representation has
strictly negative flow \(-k+\beta\lambda_GG\) near the endpoint and
terminal value zero, contradicting \(\Pi\ge0\).

If \(\beta'=\beta_{S,0}\), the same representation and discounted
survival bounded by one give, for a constant \(C>0\) and all sufficiently
nearby \(\beta>\beta'\),
\[
0\le\Pi(\beta)\le
C(\beta-\beta')\sigma(\beta)
+C\sup_{\beta'<\xi\le\beta}|V(\xi)-F|
\,[x(\beta)-x(\beta')]
=o(\beta-\beta').
\]
Here \(\int_0^{\sigma(\beta)}(\lambda_G-\lambda_B)dt
=x(\beta)-x(\beta')\), and log-odds are locally Lipschitz at \(\beta'\).
But \(\Pi_S(\beta')=0\) and \(\Pi_S'=\lambda_SG/(r+\lambda_S)>0\),
contradicting \(\Pi\ge\Pi_S\).

Now suppose that a fail-learning component ends in hiring at its upper
endpoint \(\beta''\). Matching and the lower envelope imply
\(\Pi-h\to0\), \(V\to H\), and \(\beta''\ge\beta_{S,h}\), while
\[
\lambda_B\le\lambda_S+
\frac{(\lambda_B-\lambda_G)(H-V)}{H-F},
\qquad \limsup_{\beta\uparrow\beta''}\lambda_B\le\lambda_S.
\]
Subtracting the payoff of immediate hiring from the principal's pathwise
representation gives
\[
(\Pi-h)(\beta)=
-\int_0^{\sigma(\beta)}D_P(t,\beta)
\bigl[rh(\beta_t)+k+(1-\beta_t)\lambda_B(\beta_t)B\bigr]dt.
\]
At \(\lambda_B=\lambda_S\), the bracket is affine and strictly increasing
in \(\beta\), with its zero at \(\beta_{S,h}\). Since \(B<0\), the rate
bound makes the bracket strictly positive near the endpoint if
\(\beta''>\beta_{S,h}\), contradicting \(\Pi\ge h\).

If \(\beta''=\beta_{S,h}\), the same bound instead yields, for a constant
\(C>0\) and all sufficiently nearby \(\beta<\beta''\),
\[
0\le(\Pi-h)(\beta)\le
C(\beta''-\beta)\sigma(\beta)
+C\sup_{\beta\le\xi<\beta''}|H-V(\xi)|
\,[x(\beta'')-x(\beta)]
=o(\beta''-\beta).
\]
Yet \(\Pi_S-h\) vanishes at \(\beta''\) and has strictly negative
posterior derivative \(\lambda_SG/(r+\lambda_S)-(G-B)\). Thus
\(\Pi-h<\Pi_S-h\) immediately to its left, again contradicting the lower
envelope.
\end{proof}

\subsection{Proof of Theorem~\ref{thm:psio-phase-diagram}}

\begin{proof}[Proof of Theorem~\ref{thm:psio-phase-diagram}]
The thresholds \(c_N(k)\), \(k_F\), and \(c_P(k)\) are positive, and
\(\beta^\circ\in(0,1)\).

\medskip
\noindent
\textbf{Step 1: Local experiment choice.}
On a nonstationary component, \(F<V<H\): under shirking the bad type receives
no flow payoff and a discounted terminal payoff in \([F,H]\), and bounded
rates imply positive duration and preclude either endpoint value in the
interior. In posterior log-odds \(x=\log(\beta/(1-\beta))\), write
$
\Delta:=\lambda_B-\lambda_G.
$
By Lemma~\ref{lem:extreme-test-reduction}, the chosen test is nonzero and
incentive compatibility binds:
\begin{equation}
\label{eq:phase-proof-binding}
(H-V)\lambda_G+(V-F)\lambda_B=c.
\end{equation}
Hence
\begin{equation}
\label{eq:phase-proof-rates}
\lambda_G=\frac{c-(V-F)\Delta}{H-F},
\qquad
\lambda_B=\frac{c+(H-V)\Delta}{H-F}.
\end{equation}
By Lemma~\ref{lem:bellman-characterization}, \(V\) is continuous on each
nonstationary component, while admissibility makes \(\Delta\) locally
Lipschitz. Thus \eqref{eq:phase-proof-rates} makes both rates continuous, and
every mixed point has an open mixed neighborhood. Since \(\Delta\ne0\),
\eqref{eq:bad-assessment-bellman} gives
\[
V_x=\frac{rV-\lambda_B(F-V)}{\Delta}
\quad\text{almost everywhere}.
\]
The right-hand side is continuous. Local absolute continuity of \(V\) and
the fundamental theorem of calculus therefore imply \(V\in C^1\), with
this equation holding pointwise on each nonstationary component.

On a mixed neighborhood, both nonzero extremes maximize the linear
Hamiltonian, so every point on their connecting edge does as well. In
particular, the common-rate test \((\lambda_S,\lambda_S)\) is optimal.
Its Hamiltonian is \(\lambda_S(\beta G-\Pi)\), so the Bellman equation gives
\begin{equation}
\label{eq:phase-proof-PiS-mixed}
\Pi=\Pi_S,
\end{equation}
and, since \(\Pi_x=(\Pi_S)_x\) there, the equality of marginal returns becomes
\begin{equation}
\label{eq:phase-proof-Gamma-zero}
\Gamma(\beta,V)=0,
\qquad
\Gamma(\beta,V)
:=
\beta(rG+k)(V-F)-(1-\beta)k(H-V).
\end{equation}
Equivalently,
\begin{equation}
\label{eq:phase-proof-Vtilde}
V(\beta)
=
\frac{\beta(rG+k)F+(1-\beta)kH}{rG\beta+k}.
\end{equation}
Since
$
\Gamma_\beta=rG(V-F)+k(H-F)>0,
\,
\Gamma_V=rG\beta+k>0,
\,
\Gamma(\beta^\circ,V_S)=0,
$
the mixed locus satisfies
\begin{equation}
\label{eq:phase-proof-mixed-value-side}
V>V_S \Longleftrightarrow \beta<\beta^\circ,
\qquad
V<V_S \Longleftrightarrow \beta>\beta^\circ.
\end{equation}
Differentiating \(\Gamma=0\) and using
\eqref{eq:bad-assessment-bellman} gives
$
V_x=-\frac{(H-V)(V-F)}{H-F}
$
and
\begin{equation}
\label{eq:phase-proof-singular-drift}
\Delta
=
-\frac{[r(H-F)+c](V-V_S)}{2(H-V)(V-F)}.
\end{equation}
Therefore mixed pass-learning points lie below \(\beta^\circ\), and mixed
fail-learning points lie above it.

At nonmixed points, \eqref{eq:phase-proof-binding} selects the corresponding
pure region. In a fail-only region, the active test and value equations are
\begin{equation}\label{eq:fail-region-rates}
\lambda_G(\beta)=0,
\qquad
\lambda_B(\beta)=\frac{c}{V(\beta)-F},
\end{equation}
and
\begin{equation}\label{eq:fail-region-odes}
V_x=\frac{(V-F)(rV+c)}{c},
\qquad
\Pi_x=\frac{(V-F)(r\Pi+k)}{c}+(1-\beta)\Pi.
\end{equation}
In a pass-only region, they are
\begin{equation}\label{eq:pass-region-rates}
\lambda_G(\beta)=\frac{c}{H-V(\beta)},
\qquad
\lambda_B(\beta)=0,
\end{equation}
and
\begin{equation}\label{eq:pass-region-odes}
V_x=-\frac{(H-V)rV}{c},
\qquad
\Pi_x=-\frac{(H-V)(r\Pi+k)}{c}+\beta(G-\Pi).
\end{equation}
Define the Hamiltonians at the pass-only and fail-only nonzero extremes by
\[
\Phi_P(\beta;V,\Pi)
:=
\Psi_P\left(\beta,\frac{c}{H-V(\beta)},0;\Pi\right),
\qquad
\Phi_F(\beta;V,\Pi)
:=
\Psi_P\left(\beta,0,\frac{c}{V(\beta)-F};\Pi\right).
\]
Suppressing their arguments, direct substitution yields
\begin{align}
\Phi_P-\Phi_F
&=
\frac{(H-F)(r+\lambda_S)}{V-F}(\Pi-\Pi_S)
&&\text{in a pass-only region},
\label{eq:phase-proof-gap-pass}\\
\Phi_F-\Phi_P
&=
\frac{(H-F)(r+\lambda_S)}{H-V}(\Pi-\Pi_S)
&&\text{in a fail-only region}.
\label{eq:phase-proof-gap-fail}
\end{align}

\medskip
\noindent
\textbf{Step 2: Every nonstationary component attaches to a stationary
belief.}
For a pass-learning component, the no-news direction points to its lower
endpoint. At pure points,
$
-\Delta=\frac{c}{H-V}>\lambda_S.
$
At mixed points, a failure of \(-\Delta\) to be bounded away from zero would,
by \eqref{eq:phase-proof-Vtilde} and
\eqref{eq:phase-proof-singular-drift}, force
\((\beta,V)\to(\beta^\circ,V_S)\). This cannot occur at a lower endpoint
equal to \(\beta^\circ\): mixed pass-learning points satisfy
\(\beta<\beta^\circ\), whereas a pass-learning component approaches its
lower endpoint through beliefs strictly to its right. Thus
\(-\Delta\) is bounded away from zero near every admissible interior lower
endpoint, which is reached in finite time. It cannot be a firing endpoint by
Lemma~\ref{lem:learning-phases-wrong-stopping-action}. At a hiring endpoint,
value matching gives \(V\to H\); then \(\Gamma(\beta,H)>0\) excludes mixed
points, while a pure region has \(\lambda_G=c/(H-V)\to\infty\), violating
admissibility. The endpoint cannot be zero. Mixed points cannot accumulate there:
\eqref{eq:phase-proof-Vtilde} gives \(V\to H\), and
\eqref{eq:phase-proof-singular-drift} then gives
\(|\Delta|\to\infty\). Thus a sufficiently far left tail is pure pass-only.
With \(R:=V/(H-V)\), \eqref{eq:pass-region-odes} gives
$
R_x=-\frac{rH}{c}R,
\,
R(x)=R(x_0)e^{-rH(x-x_0)/c}.
$
A positive solution has \(V\to H\) as \(x\to-\infty\) and hence an unbounded
pass rate; a negative solution reaches \(V=F\) at finite \(x\). Hence the only
solution extending to \(x\to-\infty\) with bounded rates is \(V\equiv0\). In
that case, with \(W:=\Pi+k/r\),
\[
W_x=-\left(\frac{Hr}{c}+\beta\right)W
+\beta\left(G+\frac{k}{r}\right).
\]
Boundedness and variation of constants give \(W\to0\) as \(x\to-\infty\), so
\(\Pi\to-k/r<0\), contrary to the lower envelope. Thus the lower endpoint is
stationary.

For a fail-learning component, the no-news direction points to its upper
endpoint and, at pure points,
$
\Delta=\frac{c}{V-F}>\lambda_S.
$
If \(\Delta\) were not bounded away from zero along mixed points, the mixed
locus would again force
\((\beta,V)\to(\beta^\circ,V_S)\). But mixed fail-learning points satisfy
\(\beta>\beta^\circ\), whereas a fail-learning component approaches its upper
endpoint through beliefs strictly to its left. Hence mixed fail points cannot
accumulate at an upper endpoint equal to \(\beta^\circ\), and every
admissible interior upper endpoint is reached in finite time.
Hiring is excluded by
Lemma~\ref{lem:learning-phases-wrong-stopping-action}; firing would give
\(V\to F\), which excludes mixed points and makes
\(\lambda_B=c/(V-F)\) unbounded. The endpoint cannot be one. Mixed points cannot accumulate there:
\eqref{eq:phase-proof-Vtilde} gives \(V\to F\), and
\eqref{eq:phase-proof-singular-drift} gives
\(|\Delta|\to\infty\). Thus a sufficiently far right tail is pure fail-only.
With \(Y:=V-F\), \eqref{eq:fail-region-odes} gives
$
Y_x=\frac{Y(c+rF+rY)}{c}.
$
The scalar phase portrait shows that a solution extending to
\(x\to\infty\) while remaining in \(F<V<H\) can only have
\(V\to F\) or, when \(c+rF<0\), \(V\to-c/r\). The former makes
\(\lambda_B=c/(V-F)\) unbounded. In the latter case, with
\(W:=\Pi+k/r\),
$
W_x=
\left[\frac{r(V-F)}{c}+1-\beta\right]W
-(1-\beta)\frac{k}{r}.
$
The coefficient converges to \(-(c+rF)/c>0\), so boundedness and variation of
constants give \(W\to0\). Hence \(\Pi\to-k/r<h(1)\), contrary to the lower
envelope. Thus the upper endpoint is stationary.

\medskip
\noindent
\textbf{Step 3: Stationary feasibility.}
At a stationary belief,
$
V=V_S,\, \Pi=\Pi_S(\beta).
$
The lower envelope is equivalent to
\(\beta\in[\beta_{S,0},\beta_{S,h}]\), and
\begin{equation}
\label{eq:phase-proof-stationary-width}
\beta_{S,h}-\beta_{S,0}
=
\frac{[r(H-F)+c](-B)[c-c_N(k)]}
{c\{rG(H-F)-B[r(H-F)+c]\}}.
\end{equation}
Thus the set of feasible stationary beliefs is empty if \(c<c_N(k)\), a singleton
\(\{\beta^*\}\) if \(c=c_N(k)\), and a nondegenerate interval if
\(c>c_N(k)\).

If \(c<c_N(k)\), Step 2 excludes nonstationary probation, so
\(\Pi=\max\{0,h\}\). At equality, no component can attach to \(\beta^*\).
Mixed points are excluded by \(\Gamma(\beta^*,V_S)\ne0\), or, when
\(\beta^*=\beta^\circ\), by the mixed-point side restrictions from Step 1.
A pure pass region to the right would satisfy
$
(\Pi-h)_x(\beta^*+)=-\frac{(H-V_S)k}{c}<0,
$
and a pure fail region to the left would satisfy
$
\Pi_x(\beta^*-) = \frac{(V_S-F)k}{c}>0;
$
both violate the lower envelope. Hence only the isolated stationary
convention, payoff-equivalent for the principal, remains.

Suppose \(c>c_N(k)\). Then
$
0<\beta_{S,0}<\beta^*<\beta_{S,h}<1,
\,
\Pi_S>\max\{0,h\}
\,\text{on }(\beta_{S,0},\beta_{S,h}).
$
In particular, probation is required at \(\beta^*\), so Step 2 implies the
existence of a stationary belief.

\medskip
\noindent
\textbf{Step 4: Global region configuration.}
Let \(s\) be a stationary attachment and set \(Q:=\Pi-\Pi_S\). Mixed points
cannot approach \(s\): continuity would force \(s=\beta^\circ\), but a mixed
pass component lies to the right of \(s\) and must be below
\(\beta^\circ\); the fail case is symmetric. Thus each region is pure near
its attachment.

Direct substitution gives, at every pure point,
\begin{align}
Q_x
&=
\frac{\Gamma(\beta,V)}{r(H-F)+c}
-\left[\frac{r(H-V)}{c}+\beta\right]Q
&&\text{in a pass-only region},
\label{eq:phase-proof-Q-pass}\\
Q_x
&=
\frac{\Gamma(\beta,V)}{r(H-F)+c}
+\left[\frac{r(V-F)}{c}+1-\beta\right]Q
&&\text{in a fail-only region}.
\label{eq:phase-proof-Q-fail}
\end{align}
At the stationary attachment, \(Q=0\), so the sign of the limiting
\(Q_x\) is the sign of \(\Gamma(s,V_S)\), hence the sign of
\(s-\beta^\circ\). Since \(Q\ge0\), a pass region requires
\(s\ge\beta^\circ\), and a fail region requires \(s\le\beta^\circ\).
The mixed-point side restrictions in Step 1 then exclude mixed points on the
entire component.

In the resulting pure pass region, \(V_S<V<0\), so \(V_x>0\). In the
pure fail region, \(V>\max\{F,-c/r\}\) and \(V_x>0\): the initial value
satisfies \(V_S>-c/r\), and uniqueness prevents crossing a stationary
solution of the scalar value equation. Consequently,
\[
\frac{d}{dx}\Gamma(\beta(x),V(x))
=\Gamma_\beta\beta(1-\beta)+\Gamma_VV_x>0.
\]
Apply the integrating-factor formula to
\eqref{eq:phase-proof-Q-pass}--\eqref{eq:phase-proof-Q-fail}, starting from
\(Q(s)=0\). In a pass region, the forcing term is strictly positive to the
right of the attachment. In a fail region it is strictly negative to the
left, and the integration direction is reversed. Thus, in both cases,
\begin{equation}\label{eq:phase-proof-Q-strict}
\Pi>\Pi_S\qquad\text{away from the stationary attachment}.
\end{equation}
The same integral remains strictly positive at any finite opposite endpoint:
its coefficients are bounded there because \(V\in[F,H]\), and its integrand has
a strict sign on a nonempty interval. This argument also covers
\(s=\beta^\circ\), without a separate second-derivative calculation.

There is only one stationary belief. If two existed, choose two consecutive
ones \(s_1<s_2\). Since both lie in the feasible stationary interval,
\(\Pi_S>\max\{0,h\}\) throughout \((s_1,s_2)\). The lower-envelope condition
therefore forces every intermediate belief to be on probation. By
consecutiveness, none is stationary, so \((s_1,s_2)\) is a nonstationary
component with \(Q=0\) at both endpoints, contrary to the no-return argument.

Let \(s\) denote the unique stationary belief. A sole right component must
attach at \(s=\beta_{S,0}\). Indeed, stationary feasibility gives
\(s\ge\beta_{S,0}\); if the inequality were strict, then \(\Pi_S\) would
strictly dominate stopping immediately to the left of \(s\), forcing an
additional left probation component. Its attachment condition is therefore
\(\beta_{S,0}\ge\beta^\circ\). Symmetrically, a sole left component must
attach at \(s=\beta_{S,h}\), with
\(\beta_{S,h}\le\beta^\circ\). If both components exist, then
$
s=\beta^\circ,\,
\beta_{S,0}<\beta^\circ<\beta_{S,h};
$
equality at either endpoint would violate the same derivative lower-envelope
conditions used in Step 3.

The comparisons are
\begin{align}
\beta_{S,0}-\beta^\circ
&=
\frac{k[r(H-F)+c]}
{c\{k(Hr+c)-rF(Gr+k)\}}
[c_P(k)-c],
\label{eq:phase-proof-pass-comparison}\\
\beta^\circ-\beta_{S,h}
&=
\frac{(H-F)(Gr+k)[r(H-F)+c]}
{\{k(Hr+c)-rF(Gr+k)\}
 \{rG(H-F)-B[r(H-F)+c]\}}
[k-k_F].
\label{eq:phase-proof-fail-comparison}
\end{align}
Their prefactors are positive. Finally,
\begin{equation}
\label{eq:phase-proof-threshold-order}
c_P(k)-c_N(k)=\frac{(H-F)(k-k_F)}{B}.
\end{equation}
Since \(B<0\), these signs give the four regions in the theorem. At
\(c=c_P(k)\) the fail region has zero length; at \(k=k_F\) the pass region
has zero length.

\medskip
\noindent
\textbf{Step 5: Construction and uniqueness.}
The region-specific systems are locally Lipschitz on \(F<V<H\); the stationary
attachment and \(V=V_S,\Pi=\Pi_S\) therefore determine the solution in each region uniquely.
The strict inequality \eqref{eq:phase-proof-Q-strict} and the gap formulas
\eqref{eq:phase-proof-gap-pass}--\eqref{eq:phase-proof-gap-fail} verify
the selected nonzero extreme relative to the other nonzero extreme.

The solution in a pass-only region reaches a unique hiring threshold. Continuing this solution,
\(V\uparrow0\) and
$
\Pi(\beta)\longrightarrow\frac{cG-Hk}{c+Hr}<G
\,(\beta\uparrow1).
$
Since \(\Pi-h>0\) at the attachment, a crossing exists. At any crossing
\(\Pi=h\), \(h=\Pi>\Pi_S\ge0\), and
$
(\Pi-h)_x=-\frac{(H-V)(rh+k)}{c}<0.
$
Thus every zero is crossed from positive to negative, so the crossing is
unique. Since \(\Pi>\Pi_S\ge h\) for \(\beta\le\beta_{S,h}\), it lies above
\(\beta_{S,h}\).

The solution in a fail-only region reaches a unique firing threshold. Every regime requiring
such a region has \(c>-rF\). Indeed, in the fail--pass region
\(c>c_P(k)>-rF\), while in the fail-only region
$
c>c_N(k)\ge c_N(k_F)
=-rF\frac{G-B}{G}>-rF.$
With \(Y:=V-F\) and \(Z:=\Pi/Y\),
$
Y_x=\frac{Y(c+rF+rY)}{c},
\,
Z_x=\frac{k}{c}-\left(\beta+\frac{rF}{c}\right)Z.
$

As \(x\to-\infty\),
\[
Y\sim C_0e^{(c+rF)x/c}\quad\text{for some }C_0>0,
\qquad
\beta=o(Y),
\qquad
Z\longrightarrow\frac{k}{rF}<0,
\]
where the last limit follows from variation of constants in the linear
\(Z\)-equation. Thus \(\Pi<0\) near zero, while \(\Pi>0\) at the attachment,
so a crossing exists. At every zero,
$
\Pi_x=\frac{(V-F)k}{c}>0.
$
Thus every zero is crossed from negative to positive, so the crossing is
unique. Since \(\Pi>\Pi_S\ge0\) for \(\beta\ge\beta_{S,0}\), it lies below
\(\beta_{S,0}\).

For \(c\le c_N(k)\), stop immediately, with either stopping action at
\(\beta^*\) and the optional isolated stationary convention at equality.
For \(c>c_N(k)\), attach the regions identified in Step 4, using
\(s=\beta_{S,h}\), \(s=\beta_{S,0}\), or \(s=\beta^\circ\) in the
fail-only, pass-only, or fail--pass case, respectively. Terminate each outer
region at its unique stopping crossing. A region of zero length at a regime
boundary is simply omitted.

The probation set consists of at most two finite open intervals and one
stationary point. On each interval the pure rate is smooth and bounded, and
the drift has constant nonzero sign, so the assessment is admissible. Both
values match \((V_S,\Pi_S)\) at the stationary no-news attachment. At the
other endpoint the region ends at the unique principal-indifference crossing,
which makes \(\Pi\) continuous; no matching of \(V\) is required there because
the no-news path moves away from that endpoint.

The constructed functions are bounded, satisfy the Bellman equations, and
match the prescribed values at every endpoint reached in finite no-news time.
Lemma~\ref{lem:bellman-characterization} therefore identifies them with the
induced continuation values.

When \(c\le c_N(k)\), stationary feasibility gives
\(\Pi_S\le\max\{0,h\}\), so immediate stopping satisfies the lower envelope.
On a pass-only side, \(\Pi>\Pi_S\ge0\) and \(\Pi>h\) up to the unique hiring
crossing; above it, \(h\ge\max\{0,\Pi_S\}\) because the crossing lies above
\(\beta_{S,h}\). In a sole pass equilibrium, below
\(s=\beta_{S,0}\) both \(h\) and \(\Pi_S\) are nonpositive. On a fail-only
side, the firing crossing lies below \(\beta_{S,0}\); below it both \(h\) and
\(\Pi_S\) are negative, while between it and the attachment \(\Pi>0\), and
whenever \(h\ge0\) one has \(\Pi>\Pi_S\ge h\). Above a sole fail attachment,
\(h\ge\max\{0,\Pi_S\}\). The two-sided case combines the two one-sided
arguments. Thus the lower envelope holds globally.

In every nonstationary region, the value functions are differentiable and the
Bellman equations hold. The binding incentive constraint, the strict inequality
\eqref{eq:phase-proof-Q-strict}, and the gap formulas select the prescribed
nonzero extreme; its Hamiltonian is \(r\Pi+k>0\), whereas the zero test has
Hamiltonian zero. Thus the prescribed experiment is the pointwise maximizer
over the full incentive-compatible triangle. The stationary test is prescribed
by construction. Finally, the bad type's Bellman equation, binding incentive
constraint, and stationary equality permit application of
Lemma~\ref{lem:global-effort-verification}. Hence the constructed assessments
satisfy Definition~\ref{def:principal-smooth-interior-optimal-mpbe}, and the
selected shirking behavior is optimal against every progressively measurable
effort deviation.

Steps 2–4 show that the closure of each probation component has two endpoints strictly inside \((0,1)\). Since there are finitely many such components and stationary points, their union has positive distance from \(\{0,1\}\); hence choose \(\varepsilon>0\) accordingly. Local
uniqueness of the ODEs and the crossing derivatives then give uniqueness,
modulo payoff-irrelevant choices at stopping indifference beliefs, off-path
tests, and the isolated stationary convention at \(c=c_N(k)\).
\end{proof}

\subsection{Proofs of results in Section \ref{sec:no-manipulation}}
\label{app:proof-no-manipulation}

\begin{proof}[Proof of Proposition~\ref{prop:no-manipulation-attention}]
Let \(W\) denote the principal's value in this benchmark, and let
\(m(\beta):=\max\{0,h(\beta)\}\) be the stopping payoff. We apply
\citet[Theorem~1]{che2019optimal}, using the payoff and flow-cost normalization
in Section~\ref{sec:no-manipulation} and identifying their states with the good
and bad types and their terminal actions with hiring and firing. Hiring
pays \(G\) in the good state and \(B\) in the bad state, whereas firing
pays zero in both states. The two sources provide conclusive pass and fail
news, respectively, with attention capacity \(\bar\lambda\).

\medskip
\noindent
\textbf{Step 1: Local experiment choice.}
In a continuation region with smooth values, the experiment's contribution to the
principal's Bellman equation is
\[
\beta\lambda_G(G-W)-(1-\beta)\lambda_BW
+\beta(1-\beta)(\lambda_B-\lambda_G)W'.
\]
Its maximum equals \(rW+k>0\), since \(W\geq m\geq0\). A maximizing
test is therefore nonzero and exhausts the capacity: otherwise, scaling
up both rates would strictly increase this expression. Write
\(\lambda_G=a\bar\lambda\) and
\(\lambda_B=(1-a)\bar\lambda\), where \(a\in[0,1]\). The variational
inequality is
\begin{equation}
\begin{split}
0=\max\Bigl\{
&m(\beta)-W(\beta),\\
&-k-rW(\beta)+\max_{a\in[0,1]}\bar\lambda
\bigl[a\beta\{G-W(\beta)\}-(1-a)(1-\beta)W(\beta)\\
&\hspace{43mm}+(1-2a)\beta(1-\beta)W'(\beta)\bigr]
\Bigr\}.
\end{split}
\label{eq:no-manipulation-hjb}
\end{equation}
At nonsmooth beliefs, this equation is interpreted in the viscosity sense.
The Hamiltonian is affine in \(a\), so a pure test maximizes whenever its
two endpoint values differ. The endpoints \(a=1\) and \(a=0\) are
pass-only and fail-only monitoring, with rates \((\bar\lambda,0)\) and
\((0,\bar\lambda)\), respectively. In the absence of news,
$
\dot\beta=(1-2a)\bar\lambda\beta(1-\beta).
$
Thus pass-only monitoring lowers the posterior, whereas fail-only
monitoring raises it.

\medskip
\noindent
\textbf{Step 2: Stopping.}
At the stopping payoff \(W=m\), the maximal continuation gain is
\[
\begin{cases}
\bar\lambda\beta G-k,
&\beta<\beta^*,\\
-\bar\lambda(1-\beta)B-rh(\beta)-k,
&\beta>\beta^*.
\end{cases}
\]
The first expression is increasing and the second is decreasing in
\(\beta\). Their common limit at \(\beta^*\) is
\(\bar\lambda\beta^*G-k\), which is also the maximal gain at
\(\beta^*\). Hence the upper probation-cost threshold is
\begin{equation}
\overline k^A
:=\bar\lambda\beta^*G
=\frac{\bar\lambda G(-B)}{G-B}.
\label{eq:attention-upper-cost}
\end{equation}
For \(k\geq\overline k^A\), the stopping payoff satisfies
\eqref{eq:no-manipulation-hjb}, and the verification argument in the cited
theorem gives immediate stopping. For \(k<\overline k^A\), continuing
briefly at \(\beta^*\) is strictly profitable.

For \(0<k<\overline k^A\), the same theorem gives a continuation interval
\((\underline\beta^A,\overline\beta^A)\), with pass-only monitoring near
its lower endpoint and fail-only monitoring near its upper endpoint.
Value matching and smooth pasting give
\[
\begin{aligned}
W(\underline\beta^A)&=0,
& W'(\underline\beta^A+)&=0,\\
W(\overline\beta^A)&=h(\overline\beta^A),
& W'(\overline\beta^A-)&=G-B.
\end{aligned}
\]
Evaluating the corresponding region-specific equations at the endpoints yields
$
k=\bar\lambda\underline\beta^A G,
\,
k+rh(\overline\beta^A)
=-\bar\lambda(1-\overline\beta^A)B.
$
Consequently,
\begin{equation}
\underline\beta^A=\frac{k}{\bar\lambda G},
\qquad
\overline\beta^A
=\frac{-k-(r+\bar\lambda)B}{r(G-B)-\bar\lambda B}.
\label{eq:attention-stopping-boundaries}
\end{equation}
For \(0<k<\overline k^A\), these boundaries satisfy
\(0<\underline\beta^A<\beta^*<\overline\beta^A<1\).

\medskip
\noindent
\textbf{Step 3: Interior switches.}
A stationary continuation policy uses
\(\lambda_G=\lambda_B=\bar\lambda/2\). Its payoff is
$
W_S^A(\beta):=\frac{\bar\lambda\beta G-2k}{2r+\bar\lambda}.
$
In the low-cost configuration of the cited theorem, the value functions in the two inner
regions attach smoothly to this stationary payoff. At their attachment,
\(W=W_S^A\) and \(W'=(W_S^A)'\), and the two pure-test Hamiltonians
coincide. Their difference, evaluated at \(W_S^A\), is
\[
\bar\lambda\bigl[\beta G+(1-2\beta)W_S^A
-2\beta(1-\beta)(W_S^A)'\bigr]
=
\frac{2\bar\lambda}{2r+\bar\lambda}
\bigl[\beta(rG+2k)-k\bigr].
\]
The stationary attachment is therefore
\begin{equation}
\beta_o^A=\frac{k}{rG+2k}.
\label{eq:attention-stationary-belief}
\end{equation}

The lower probation-cost threshold in the cited theorem specializes to
\begin{align}
D_A
&:=1+\left(\frac{2r+\bar\lambda}{\bar\lambda}\right)^{\bar\lambda/r},
\nonumber\\
\underline k^A
&:=\max\left\{0,
\min\left\{
\overline k^A,
\frac{(r+\bar\lambda)G}{D_A}-rG,
\frac{(r+\bar\lambda)(-B)}{D_A}
\right\}\right\}.
\label{eq:attention-lower-cost}
\end{align}
In particular, \(0\leq\underline k^A\leq\overline k^A\).

If \(\underline k^A\leq k<\overline k^A\), the theorem's crossing
argument gives a switching belief
\(\check\beta^A\in(\underline\beta^A,\overline\beta^A)\), with
pass-only monitoring below it and fail-only monitoring above it. By the
drift formula in Step~1, no news moves the posterior toward the lower
stopping boundary in the pass-only region and toward the upper stopping
boundary in the fail-only region. This proves \textnormal{(ii)}.

If \(0<k<\underline k^A\), the same argument gives cutoffs
$
\underline\beta^A<\beta_-^A<\beta_o^A<\beta_+^A<\overline\beta^A.
$
The two outer regions use pass-only and fail-only monitoring,
respectively, whereas the two inner regions use fail-only and pass-only
monitoring, respectively, and attach to the stationary policy at
\(\beta_o^A\). Thus no news moves outward in the outer regions and
inward in the inner regions, as claimed in \textnormal{(iii)}.

At each repelling switch, choose an optimal outgoing no-news trajectory,
as allowed in Section~\ref{sec:no-manipulation}. The cited verification
argument establishes optimality over the benchmark's history-adapted
attention controls and stopping rules. Together with Step~2, this proves
all three cases.
\end{proof}

\subsection{Proofs of results in Section \ref{sec:commitment}}
\label{sec:commitment-appendix}

\begin{proof}[Proof of Theorem~\ref{thm:optimal-commitment-bounded}]
Define
\begin{align}
\Pi^{\mathsf P}(\beta)
&:=
\beta\frac{cG-kH}{c+rH}
-(1-\beta)\frac{k}{r},                                     
\label{eq:bounded-commitment-pass-payoff}\\
\Pi^{\mathsf M}(\beta)
&:=
\beta\frac{G(c+rF)-k(H-F)}{c+rH}.
\label{eq:bounded-commitment-screening-coefficient}
\end{align}
The first expression is the payoff from policy \textnormal{(iv)}. The second
is an auxiliary coefficient used to bound the payoff from screening.

When \(c+rF>0\), define
\begin{equation}
\lambda_G^{\bar\lambda}
:=
\frac{cr+\bar\lambda(c+rF)}
     {rH+\bar\lambda(H-F)}
\label{eq:bounded-commitment-pass-rate}
\end{equation}
and let
\begin{equation}
\Pi_{\bar\lambda}^{\mathsf M}(\beta)
:=
\beta
\frac{\lambda_G^{\bar\lambda}G-k}
     {r+\lambda_G^{\bar\lambda}}
-(1-\beta)\frac{k}{r+\bar\lambda}
\label{eq:bounded-commitment-maximal-fail-payoff}
\end{equation}
be the payoff from policy \textnormal{(iii)}. We prove that the principal's
optimal commitment payoff is
\begin{equation}
\Pi_{\bar\lambda}^{C}(\beta)
=
\begin{cases}
\displaystyle
\max\left\{
0,\,
h(\beta),\,
\Pi^{\mathsf P}(\beta),\,
\Pi_{\bar\lambda}^{\mathsf M}(\beta)
\right\},
& c+rF>0,\\[2mm]
\displaystyle
\max\left\{
0,\,
h(\beta),\,
\Pi^{\mathsf P}(\beta)
\right\},
& c+rF\leq0.
\end{cases}
\label{eq:bounded-commitment-value}
\end{equation}

Fix an arbitrary feasible commitment policy and a best response of the bad
type. For each \(\theta\in\{G,B\}\), let \(\tau_\theta\in[0,\infty]\) be the
terminal time, where \(\tau_\theta=\infty\) if probation never ends. When
\(\tau_\theta<\infty\), let
\(a_\theta\in\{\mathsf h,\mathsf f\}\) denote the terminal action. Define
\begin{align}
X_\theta
&:=
\mathbb E_\theta\left[
e^{-r\tau_\theta}
\mathbf 1_{\{a_\theta=\mathsf h\}}
\right],\\
Z_\theta
&:=
\mathbb E_\theta\left[
e^{-r\tau_\theta}
\mathbf 1_{\{a_\theta=\mathsf f\}}
\right],\\
Q_\theta
&:=
\mathbb E_\theta\left[
\int_0^{\tau_\theta}e^{-rt}\,dt
\right].
\end{align}
We adopt the convention \(e^{-r\infty}=0\). Since every finite terminal
history ends in either hiring or firing,
\begin{equation}
X_\theta+Z_\theta+rQ_\theta=1,
\qquad
\theta\in\{G,B\}.
\label{eq:bounded-commitment-accounting}
\end{equation}

\medskip
\noindent
\textit{Monotonicity and the implication of the rate bound.}
The discounted terminal allocations satisfy
\begin{equation}
X_G\geq X_B,
\qquad
Z_B\geq Z_G.
\label{eq:bounded-commitment-monotonicity}
\end{equation}
Indeed, couple the two types along their common no-news history. Construct
pass arrivals from a common Poisson random measure and obtain the bad type's
pass process by predictable thinning with retention probability equal to his
effort. Thus every pass of the bad type is also a pass of the good type. Add
a fail process only for the bad type. Under this coupling, whenever the bad
type is hired, the good type is hired weakly earlier, and whenever the good
type is fired, the bad type is fired weakly earlier. This gives
\eqref{eq:bounded-commitment-monotonicity} after discounting and taking
expectations.

The bound on test intensities further implies
\begin{equation}
Z_B-Z_G\leq\bar\lambda Q_B.
\label{eq:bounded-commitment-rate-restriction}
\end{equation}
To verify this inequality, condition on all public and private randomization.
Along the resulting no-news path, let \(T\) be the time at which the
principal stops in the absence of a signal, and let
\(A\in\{\mathsf h,\mathsf f\}\) be the corresponding action when \(T<\infty\).
Write \(g_t\), \(b_t\), and \(e_t\) for the pass intensity, fail intensity,
and bad type's effort along this path. Define the discounted no-signal
probabilities
\[
D_G(t)
:=
\exp\left\{
-rt-\int_0^t g_s\,ds
\right\},
\qquad
D_B(t)
:=
\exp\left\{
-rt-\int_0^t
\bigl[e_sg_s+(1-e_s)b_s\bigr]\,ds
\right\}.
\]
Conditional on the fixed randomization,
$
Q_B=\int_0^T D_B(t)\,dt
$
and
\begin{equation}
Z_B-Z_G
=
\int_0^T D_B(t)(1-e_t)b_t\,dt
+
\mathbf 1_{\{A=\mathsf f\}}
\bigl[D_B(T)-D_G(T)\bigr].
\label{eq:bounded-commitment-firing-gap}
\end{equation}
When \(T=\infty\), the terminal term is zero and the desired rate bound
follows directly from \((1-e_t)b_t\le\bar\lambda\). For \(T<\infty\), let
$
\Delta
:=
\int_0^T(1-e_t)(g_t-b_t)\,dt.
$
Since \(D_G(T)=D_B(T)e^{-\Delta}\), the inequality
\(1-e^{-z}\leq z_+\), where \(z_+:=\max\{z,0\}\), and the monotonicity of
\(D_B\) imply
\begin{align*}
D_B(T)-D_G(T)
&\leq
D_B(T)\Delta_+\\
&\leq
\int_0^T
D_B(t)(1-e_t)(g_t-b_t)_+\,dt.
\end{align*}
Substituting this bound into
\eqref{eq:bounded-commitment-firing-gap} gives
\begin{align*}
Z_B-Z_G
&\leq
\int_0^T
D_B(t)(1-e_t)
\left[b_t+(g_t-b_t)_+\right]dt\\
&=
\int_0^T
D_B(t)(1-e_t)\max\{g_t,b_t\}\,dt\\
&\leq
\bar\lambda\int_0^T D_B(t)\,dt
=
\bar\lambda Q_B.
\end{align*}
Averaging over the conditioned randomization proves
\eqref{eq:bounded-commitment-rate-restriction}.

\medskip
\noindent
\textit{The full-imitation constraint.}
By exerting high effort forever, the bad type can generate the good type's
public history distribution and receive \(HX_G+FZ_G-cQ_G\). His actual
effort costs are nonnegative. A necessary incentive constraint is therefore
\begin{equation}
HX_B+FZ_B\ge HX_G+FZ_G-cQ_G.
\label{eq:bounded-commitment-relaxed-ic}
\end{equation}

Set
$
x:=X_B,
\,
y:=rQ_B,
\,
u:=X_G-X_B,
\,
w:=Z_B-Z_G.
$
By \eqref{eq:bounded-commitment-monotonicity},
\(x,y,u,w\geq0\). The accounting identities imply
$
Z_B=1-x-y,
\,
Z_G=1-x-y-w,
\,
rQ_G=y+w-u.
$
Consequently,
\begin{equation}
x+y+w=1-Z_G\leq1
\label{eq:bounded-commitment-resource}
\end{equation}
and, by \eqref{eq:bounded-commitment-rate-restriction},
\begin{equation}
w\leq\frac{\bar\lambda}{r}y.
\label{eq:bounded-commitment-linear-rate-restriction}
\end{equation}
Substituting the accounting identities into
\eqref{eq:bounded-commitment-relaxed-ic} yields
\begin{equation}
(c+rH)u
\leq
cy+(c+rF)w.
\label{eq:bounded-commitment-linear-ic}
\end{equation}

\medskip
\noindent
\textit{Upper bound.}
The principal's payoff under the arbitrary commitment is
\begin{align}
\Pi
&=
\beta\bigl(GX_G-kQ_G\bigr)
+(1-\beta)\bigl(BX_B-kQ_B\bigr)\nonumber\\
&=
h(\beta)x
+\beta\left(G+\frac{k}{r}\right)u
-\frac{k}{r}y
-\beta\frac{k}{r}w.
\label{eq:bounded-commitment-principal-payoff}
\end{align}
Because \(\beta(G+k/r)\ge0\), inequality
\eqref{eq:bounded-commitment-linear-ic} gives
\begin{equation}
\Pi
\leq
h(\beta)x
+\Pi^{\mathsf P}(\beta)y
+\Pi^{\mathsf M}(\beta)w.
\label{eq:bounded-commitment-preliminary-bound}
\end{equation}

Suppose first that \(c+rF\leq0\). Then
$
\Pi^{\mathsf M}(\beta)
=
\beta
\frac{G(c+rF)-k(H-F)}{c+rH}
\le0.
$
Since \(w\geq0\), \eqref{eq:bounded-commitment-preliminary-bound} and
\(x+y\leq1\) imply
$
\Pi
\leq
\max\left\{
0,\,
h(\beta),\,
\Pi^{\mathsf P}(\beta)
\right\}.
$

Now suppose that \(c+rF>0\). A direct substitution of
\eqref{eq:bounded-commitment-pass-rate} into
\eqref{eq:bounded-commitment-maximal-fail-payoff} gives
\begin{equation}
\Pi_{\bar\lambda}^{\mathsf M}(\beta)
=
\frac{r}{r+\bar\lambda}\Pi^{\mathsf P}(\beta)
+
\frac{\bar\lambda}{r+\bar\lambda}\Pi^{\mathsf M}(\beta).
\label{eq:bounded-commitment-payoff-identity}
\end{equation}
Indeed,
$
r+\lambda_G^{\bar\lambda}
=
\frac{(r+\bar\lambda)(c+rH)}
     {rH+\bar\lambda(H-F)},
$
from which \eqref{eq:bounded-commitment-payoff-identity} follows after
collecting terms.

Using \eqref{eq:bounded-commitment-payoff-identity}, the right-hand side of
\eqref{eq:bounded-commitment-preliminary-bound} can be written as
\[
h(\beta)x
+
\Pi^{\mathsf P}(\beta)
\left(y-\frac{r}{\bar\lambda}w\right)
+
\Pi_{\bar\lambda}^{\mathsf M}(\beta)
\frac{r+\bar\lambda}{\bar\lambda}w.
\]
The three displayed coefficients are nonnegative by
\eqref{eq:bounded-commitment-linear-rate-restriction}, and their sum is
\(x+y+w\leq1\). Therefore,
\[
\Pi
\leq
\max\left\{
0,\,
h(\beta),\,
\Pi^{\mathsf P}(\beta),\,
\Pi_{\bar\lambda}^{\mathsf M}(\beta)
\right\}.
\]
This establishes the upper bound in \eqref{eq:bounded-commitment-value}.

\medskip
\noindent
\textit{Attainability.}
Immediate firing and immediate hiring attain \(0\) and \(h(\beta)\),
respectively.

For perpetual pass-only monitoring, set \(\lambda_G=c/H\) and
\(\lambda_B=0\), which is feasible because \(\bar\lambda\ge c/H\).
The constant bad-type value zero satisfies the controlled Bellman equation
with \(\Gamma=\lambda_GH-c=0\). The verification argument of
Lemma~\ref{lem:global-effort-verification} therefore makes every effort
process optimal. Under the maintained shirking selection, the bad type never
exits and the good type is hired at rate \(c/H\), giving
\[
\beta\frac{(c/H)G-k}{r+c/H}-(1-\beta)\frac{k}{r}
=\Pi^{\mathsf P}(\beta).
\]

When \(c+rF>0\), take \(\lambda_B=\bar\lambda\) and the pass rate
\(\lambda_G^{\bar\lambda}\) in
\eqref{eq:bounded-commitment-pass-rate}. Its numerator is positive and
\[
\frac{c}{H}-\lambda_G^{\bar\lambda}
=\frac{-\bar\lambda F(c+rH)}{H[rH+\bar\lambda(H-F)]}>0,
\]
so both rates satisfy the bound. With
\(V_B^{\bar\lambda}=\bar\lambda F/(r+\bar\lambda)\), the rate formula gives
\[
rV_B^{\bar\lambda}=\bar\lambda(F-V_B^{\bar\lambda}),
\qquad
\lambda_G^{\bar\lambda}(H-V_B^{\bar\lambda})
+\bar\lambda(V_B^{\bar\lambda}-F)=c.
\]
Thus the same verification argument again makes every effort process
optimal. Selecting shirking yields
\[
\beta\frac{\lambda_G^{\bar\lambda}G-k}{r+\lambda_G^{\bar\lambda}}
-(1-\beta)\frac{k}{r+\bar\lambda}
=\Pi_{\bar\lambda}^{\mathsf M}(\beta).
\]
All terms in \eqref{eq:bounded-commitment-value} are therefore attainable,
which completes the proof.
\end{proof}

\begin{proof}[Proof of Proposition~\ref{prop:strategic-good-type}]
See the Online Appendix.
\end{proof}

\begin{proof}[Proof of Proposition~\ref{prop:monotone-imperfect-tests}]
See the Online Appendix.
\end{proof}

\newpage

\begingroup
\renewcommand{\baselinestretch}{1.1}
\renewcommand{\bibfont}{\small\selectfont}
\setlength{\bibsep}{3pt}

\bibliographystyle{plainnat}
\bibliography{ref}
\endgroup

\ifcomparativestaics

\fi 

\newpage

\ifonlineappendix

\begin{center}
\LARGE{Dynamic Test Design}\smallskip\\
\Large{Online Appendix}\\
{\normalsize{\today}}
\end{center}
\begin{center}
\smallskip

\textbf{Doruk Cetemen and Francesc Dilm\'e}

LUISS and EIEF, and University of Bonn\\

\href{mailto:dcetemen@luiss.it}{\texttt{dcetemen@luiss.it}} and \href{mailto:fdilme@uni-bonn.de}{\texttt{fdilme@uni-bonn.de}}
\end{center}

\section{Additional extensions}
\label{sec:extensions-appendix}

\subsection{Productive probation}
\label{sec:ext-productive}

Probation may generate output as well as information. This extension asks
how the value of retaining an agent during evaluation affects whether and
how the principal screens. Suppose probation generates flow output
\(y_G\) from a good type and \(y_B\) from a bad type, so the principal's
flow payoff is \(y(\beta)-k\), where
\(y(\beta)=y_B+\beta\Delta_y\) and \(\Delta_y:=y_G-y_B\).
Output is not separately observed and supplies no information beyond the
test outcomes. All other baseline primitives are unchanged.

We maintain
\begin{equation}
-\lambda_S G<\Delta_y<rG,
\qquad k>y_B,
\label{eq:productive-maintained-restrictions}
\end{equation}
and retain
Definition~\ref{def:principal-smooth-interior-optimal-mpbe}, with the
modified flow payoff and stationary-test guarantee.

\begin{proposition}
\label{prop:productive-probation-phase-diagram}
Under \eqref{eq:productive-maintained-restrictions}, there is a unique
equilibrium up to the baseline boundary conventions. The phase diagram
retains the four-regime classification and probation-region structure of
Theorem~\ref{thm:psio-phase-diagram}, with thresholds
\begin{equation}
\begin{aligned}
k_N^y(c)&:=k_N(c)+y(\beta^*),\\
k_P^y(c)&:=k_P(c)+y_G,\\
k_F^y&:=k_F+y_B.
\end{aligned}
\label{eq:productive-cost-thresholds}
\end{equation}
Nondegenerate probation occurs if and only if \(k<k_N^y(c)\).
In this range, the equilibrium is pass-only if \(k\ge k_P^y(c)\),
fail-only if \(k\ge k_F^y\), and fail--pass otherwise. At a regime
boundary, a region of zero length is omitted; at \(k=k_N^y(c)\), the
isolated stationary convention at \(\beta^*\), payoff-equivalent for the
principal, is permitted. Some regimes may lie outside the maintained parameter domain.

If \(y_G=y_B=y\), the model is exactly the baseline with probation
cost \(k-y>0\).
\end{proposition}

Productivity leaves the bad type's incentive constraint and the stationary
common rate \(\lambda_S=c/(H-F)\) unchanged. At this rate, the principal's
payoff is
\begin{equation}
\Pi_S^y(\beta)
=\frac{y(\beta)-k+\lambda_S\beta G}{r+\lambda_S}.
\label{eq:productive-stationary-payoff}
\end{equation}
At \(\beta^*\), both stopping actions yield zero, so the stationary test
is worthwhile exactly when \(k<k_N(c)+y(\beta^*)\). Expected output at
the static cutoff therefore shifts the threshold for whether probation
occurs.

The other two shifts reflect the different costs of delaying hiring and
firing. Good-type output offsets the cost of delaying hiring, raising the
probation cost at which pass-only monitoring becomes optimal. Bad-type
output offsets the cost of delaying firing, similarly raising the fail-only
threshold. Productivity thus changes not only the value of screening but
also the direction of monitoring. Only when output is type-independent
can both effects be captured by a common reduction in the probation cost.

The restrictions in \eqref{eq:productive-maintained-restrictions} are
substantive. They imply \(y_B-k<0\) and \(y_G-k<rG\), so output alone
does not justify retaining a known type, and preserve the ordering of the
stationary payoff's crossings with stopping. If instead \(k<y_B\) or
\(y_G-k>rG\), perpetual probation without testing can dominate firing a
known bad type or hiring a known good type, respectively. Probation then
need not remain a temporary screening stage.

\subsection{Heterogeneous discount rates}
\label{sec:ext-het}

The baseline common discount rate ties together the principal's
willingness to wait for information and the bad type's incentive to
manipulate. To separate these channels, suppose the principal discounts
at rate \(r_P>0\) and the agent at rate \(r_A>0\). All other primitives
are unchanged. We retain
Definition~\ref{def:principal-smooth-interior-optimal-mpbe}, using each
player's discount rate and the corresponding stationary-test guarantee.

\begin{proposition}
\label{prop:heterogeneous-phase-diagram}
For every \(r_P,r_A>0\), there is a unique equilibrium up to the baseline
boundary conventions. The phase diagram retains the four regimes and
probation-region structure of Theorem~\ref{thm:psio-phase-diagram}.
At each cost pair \((k,c)\), the regime is that of the baseline with
common discount rate
\begin{equation}
q(r_A,r_P;c):=
\frac{r_A[c+(H-F)r_P]}{c+(H-F)r_A}.
\label{eq:heterogeneous-composite-rate}
\end{equation}
In particular, nondegenerate probation occurs if and only if
\(k<k_N(c)\).

Fix \(c,k>0\) with \(k<k_N(c)\). Holding \(r_A\) fixed, the equilibrium
is pass-only for all sufficiently large finite \(r_P\). Holding \(r_P\)
fixed, it is fail-only for all sufficiently small positive \(r_A\).
Along any sequence of positive discount rates
\((r_{A,n},r_{P,n})\to(0,0)\), it is also fail-only for all sufficiently
large \(n\).
\end{proposition}

Heterogeneous discounting therefore moves the boundaries between the
baseline regimes without introducing new regimes or changing the
no-probation boundary. The reduction to \(q\) concerns the regime
classification, not continuation values or belief thresholds.

The stationary test makes clear why the no-probation boundary is
unchanged. Its common rate remains \(\lambda_S=c/(H-F)\), while its
continuation values are
\begin{equation}
\label{eq:heterogeneous-stationary-values}
V_S=\frac{\lambda_S F}{r_A+\lambda_S},
\qquad
\Pi_S(\beta)=\frac{\lambda_S\beta G-k}{r_P+\lambda_S}.
\end{equation}
At the static cutoff, this test strictly improves on stopping exactly
when \(k<\lambda_S\beta^*G=k_N(c)\), independently of both discount
rates. Discounting changes the value of this opportunity to learn,
but not whether it is worth pursuing.

Holding \(r_A\) fixed, principal impatience leaves \(V_S\), and hence
the maximal incentive-compatible pass and fail rates at a stationary
attachment, unchanged. It instead changes the relative returns to the
two tests. A pass brings forward a positive hiring payoff and saves
probation costs, whereas a fail only saves probation costs. A more
impatient principal therefore favors pass monitoring, eventually
eliminating the fail-only region without eliminating probation.

Agent patience instead works through incentives. As \(r_A\downarrow0\),
the bad type discounts eventual firing less, so \(V_S\downarrow F\).
A failure then causes little additional loss, whereas a pass offers
a larger gain. The maximal fail rate \(c/(V_S-F)\) therefore diverges,
while the maximal pass rate \(c/(H-V_S)\) falls to \(\lambda_S\).
More intensive fail monitoring can be sustained without inducing
effort, and a sufficiently patient agent leads to fail-only monitoring.

This incentive effect also determines the regime when both players
become patient. With positive effort and probation costs fixed, joint
patience eventually selects fail-only monitoring whenever probation
is active, regardless of the relative rates at which the discount
rates vanish. The conclusion concerns a limit of positive-discount
equilibria, not the game with both discount rates set to zero.

\section{Proofs of results in Sections \ref{sec:extensions} and \ref{sec:extensions-appendix}}
\label{sec:extensions-appendix-proofs}

\begin{proof}[Proof of Proposition~\ref{prop:strategic-good-type}]
Fix a baseline equilibrium. Retain \(V\) for the bad type's value, and let
\(V_G\) be the good type's value under high effort and the same policy.
The principal's value and signal laws are unchanged if the good type indeed
chooses high effort.

The baseline incentive constraint binds everywhere on probation. Hence the
controlled Bellman equation for an agent with effort cost \(c\) has
\(\Gamma_V=0\), and Lemma~\ref{lem:global-effort-verification} implies
that always exerting high effort also yields \(V\). By
Theorem~\ref{thm:psio-phase-diagram}, no-news paths lead to a stationary
belief rather than to stopping. Under high effort there are no fails, and
the positive stationary pass rate implies eventual hiring almost surely.
For each probation belief \(\beta\), define
\[
Q_G(\beta):=
\mathbb E_\beta^{e=1}\left[
\int_0^{\tau_{\mathrm{end}}}e^{-rt}dt\right],
\qquad 0<Q_G(\beta)\le\frac1r.
\]
Discounted-time accounting then gives
\[
V=H-(rH+c)Q_G,
\qquad V_G=H-(rH+c_H)Q_G.
\]
In particular, \(F<V<V_G<H\) on probation. Consequently,
\[
\begin{aligned}
\frac{H-V_G}{c_H}-\frac{H-V}{c}
 &=\frac{c-c_H}{cc_H}\,rH Q_G>0,\\
\frac{V_G-F}{c_H}-\frac{V-F}{c}
 &=\frac{c-c_H}{cc_H}\,[H-F-rH Q_G]>0.
\end{aligned}
\]
The second inequality uses \(H-F-rH Q_G\ge-F>0\). Every nonzero test
therefore provides a strictly larger gross effort gain relative to cost for
the good type than for the bad type. On the prescribed binding test, this
makes high effort strictly optimal for the good type. The two accounting identities express \(V_G\) as an affine function of
\(V\), so it is bounded, locally absolutely continuous, and matched at
reached endpoints,
so Lemma~\ref{lem:global-effort-verification} verifies this choice against
all progressively measurable deviations.

For any pointwise testing deviation \((\ell_G,\ell_B)\), choose pure best
responses, selecting high
effort for the good type and shirking for the bad type at their respective
indifferences. The strict normalized inequalities exclude the reverse
separation in which only the bad type works. A separating test with the good
type working and the bad type shirking belongs to the baseline feasible
triangle and generates the same principal Hamiltonian. A test inducing both
types to work has net first-order gain
\[
-k-r\Pi(\beta)+\ell_G[h(\beta)-\Pi(\beta)]<0,
\]
since $\Pi(\beta)\ge\max\{0,h(\beta)\}$. If both shirk, their identical
fail laws convey no information, and the net first-order gain is
$-k-(r+\ell_B)\Pi(\beta)<0$. Both tests are therefore dominated.

Thus the prescribed separating test remains pointwise optimal. The stopping
payoffs and stationary-test guarantee are unchanged, completing the
extension of the baseline equilibrium.
\end{proof}

\begin{proof}[Proof of Proposition~\ref{prop:monotone-imperfect-tests}]
Fix an equilibrium characterized in
Theorem~\ref{thm:psio-phase-diagram}, with policy
$(a,\lambda_G,\lambda_B)$ and continuation values $(\Pi,V)$. Retain the
action rule and prescribe
\[
(\lambda_G^1(\beta),\lambda_G^0(\beta),
 \lambda_B^1(\beta),\lambda_B^0(\beta))
=
(\lambda_G(\beta),0,0,\lambda_B(\beta)),
\qquad \beta\in[0,1].
\]
Retain high effort by the good type and shirking by the bad type, including
the selection of shirking at indifference. Under this prescription, every
progressively measurable effort process generates exactly the baseline
signal intensities and payoffs. Consequently, the induced values, belief
dynamics, admissibility, and global optimality of shirking are unchanged.
We verify the principal's conditions under the enlarged menu, retaining
Definition~\ref{def:principal-smooth-interior-optimal-mpbe}.

Fix a nonstationary probation belief
$\beta\in\mathcal K\setminus\mathcal S$ and an arbitrary monotone test
$(\ell_G^1,\ell_G^0,\ell_B^1,\ell_B^0)\in\mathbb R_+^4$. Define its
informative rates by
\[
\widetilde\ell_G:=\ell_G^1-\ell_G^0\ge0,
\qquad
\widetilde\ell_B:=\ell_B^0-\ell_B^1\ge0.
\]
For every effort $e\in[0,1]$, the pass and fail intensities are
$\ell_G^0+e\widetilde\ell_G$ and
$\ell_B^1+(1-e)\widetilde\ell_B$, respectively. Thus the test consists of
the baseline test $(\widetilde\ell_G,\widetilde\ell_B)$ together with
effort-independent pass and fail hazards $\ell_G^0$ and $\ell_B^1$.
Under assessed shirking, these common hazards cancel from the difference
between the two types' total signal rates, leaving no-news drift
$\beta(1-\beta)(\widetilde\ell_B-\widetilde\ell_G)$.

By Lemma~\ref{lem:bellman-characterization}, $V$ is continuous at $\beta$.
Holding the public assessment fixed, the bad type's local payoff is affine
in effort, with coefficient
$\widetilde\ell_G[H-V(\beta)]
+\widetilde\ell_B[V(\beta)-F]-c$.
Hence shirking is incentive compatible precisely when
\[
\widetilde\ell_G[H-V(\beta)]
+\widetilde\ell_B[V(\beta)-F]\le c,
\]
or, equivalently,
$(\widetilde\ell_G,\widetilde\ell_B)\in\mathcal A(\beta;V)$.

Suppose this constraint holds. The common pass hazard replaces
continuation by hiring without changing the posterior, while the common
fail hazard replaces continuation by firing. The principal's first-order
payoff effect therefore satisfies
\[
\begin{aligned}
&\Psi_P(\beta,\widetilde\ell_G,\widetilde\ell_B;\Pi)
+\ell_G^0[h(\beta)-\Pi(\beta)]
-\ell_B^1\Pi(\beta)\\
&\qquad\le
\Psi_P(\beta,\widetilde\ell_G,\widetilde\ell_B;\Pi)
\le r\Pi(\beta)+k.
\end{aligned}
\]
The first inequality follows from
$\Pi(\beta)\ge\max\{0,h(\beta)\}$; the second follows from baseline
experiment optimality and the principal's Bellman equation. The embedded
baseline test attains the final bound, so it remains pointwise optimal
among all shirking-compatible monotone tests.

For strict dominance of positive error rates, we verify that
$\Pi(\beta)>\max\{0,h(\beta)\}$ at every nonstationary probation belief.
By Theorem~\ref{thm:psio-phase-diagram}, each nonstationary component is
one-sided and attaches to a stationary belief $s\in\mathcal S$, where
$\Pi_S(s)=\Pi(s)\ge\max\{0,h(s)\}$. On a pass-only component,
$\beta>s$ and $\Pi_S'=\lambda_SG/(r+\lambda_S)>0$, so
\[
\Pi(\beta)\ge\Pi_S(\beta)>\Pi_S(s)\ge0.
\]
On a fail-only component, $\beta<s$ and
$(\Pi_S-h)'=[\lambda_SB-r(G-B)]/(r+\lambda_S)<0$, so
\[
\Pi(\beta)-h(\beta)
\ge\Pi_S(\beta)-h(\beta)
>\Pi_S(s)-h(s)\ge0.
\]
For the remaining inequality on each component, suppose first that
$\Pi(\beta)=h(\beta)$ at a pass-only interior point. The lower envelope
and differentiability imply $\Pi'(\beta)=G-B$. Since
$G-h(\beta)=(1-\beta)(G-B)$, substitution into the pass-only Hamiltonian
gives $\Psi_P=0$. Alternatively, if $\Pi(\beta)=0$ at a fail-only interior
point, the lower envelope implies $\Pi'(\beta)=0$, again giving
$\Psi_P=0$. Both conclusions contradict the Bellman equation
$\Psi_P=r\Pi(\beta)+k>0$. Thus both stopping inequalities are strict
throughout $\mathcal K\setminus\mathcal S$. The first inequality in the
Hamiltonian comparison is consequently strict whenever
$\ell_G^0+\ell_B^1>0$.

No additional profitable pointwise deviation arises from another effort
response. At a nonstationary probation belief, a monotone test violating
the displayed incentive constraint induces high effort from the bad type.
Both types then have the same signal law, and the principal's net
first-order gain is
\[
-k-r\Pi(\beta)
+\ell_G^1[h(\beta)-\Pi(\beta)]
-\ell_B^1\Pi(\beta)<0.
\]
When the incentive constraint binds, an intermediate effort
$e\in(0,1)$ may also be a best response. Its assessed no-news drift is
$(1-e)\beta(1-\beta)(\widetilde\ell_B-\widetilde\ell_G)$, and all
signal-payoff terms are affine in $e$. Its first-order gain is therefore
the corresponding convex combination of the shirking and high-effort
gains. It is nonpositive, and strictly negative whenever an error rate
is positive.

Finally, consider the stationary guarantee under the maintained shirking
assessment. Fix any $\beta\in(0,1)$ and any shirking-compatible monotone
test that is stationary at $\beta$. Stationarity requires
$\widetilde\ell_G=\widetilde\ell_B=:\ell$, and its incentive constraint
reduces to $\ell(H-F)\le c$, or $0\le\ell\le\lambda_S$. Holding this
test fixed until a signal arrives yields payoff
\[
\frac{-k+\ell\beta G+\ell_G^0h(\beta)}
     {r+\ell+\ell_G^0+\ell_B^1}
\le
\frac{(r+\ell)\Pi_S(\beta)+\ell_G^0h(\beta)}
     {r+\ell+\ell_G^0+\ell_B^1}
\le
\max\{\Pi_S(\beta),h(\beta),0\}.
\]
For the first inequality, the error-free stationary payoff
$(\ell\beta G-k)/(r+\ell)$ is increasing in $\ell$, with derivative
$(r\beta G+k)/(r+\ell)^2>0$, and therefore does not exceed
$\Pi_S(\beta)$. For the second, the middle expression is a convex
combination of $\Pi_S(\beta)$, $h(\beta)$, and $0$, with weights
proportional to $r+\ell$, $\ell_G^0$, and $\ell_B^1$. At $\beta=0$ and
$\beta=1$, testing cannot improve on the respective stopping payoffs
$0$ and $G$, since type is known, terminal payoffs cannot exceed those
bounds, and probation is costly. The enlarged stationary menu thus adds
no lower-envelope requirement, while the original stationary test and
both stopping actions remain available.

Retaining the baseline binding stationary test and all stopping and
boundary conventions therefore satisfies the remaining equilibrium
conditions. The embedded assessment has exactly the original policy,
continuation values, and thresholds, including in the no-probation case
and under the permitted isolated stationary convention.
\end{proof}

\subsection{Proofs of the additional extensions}

\begin{proof}[Proof of Proposition~\ref{prop:productive-probation-phase-diagram}]
We adapt the proof of Theorem~\ref{thm:psio-phase-diagram}. Only the
principal's flow payoff changes; the bad type's Bellman equation and
incentive constraint retain their baseline form.

\medskip
\noindent
\textbf{Step 1: Experiment choice and stationary feasibility.}
Write \(b=\beta(1-\beta)(\lambda_B-\lambda_G)\). At nonstationary
probation beliefs, the principal's Bellman equation is
\[
r\Pi=y(\beta)-k+b\Pi'+\beta\lambda_G(G-\Pi)
                         -(1-\beta)\lambda_B\Pi.
\]
The maintained restrictions imply \(y_G-k<rG\) and \(y_B-k<0\).
Conditional on the good type, any positive period of probation is therefore
worth strictly less than \(G\); conditional on the bad type, it is worth
strictly less than zero. Hence \(\Pi<\beta G\) on probation. A feasible
positive common-rate test has positive Hamiltonian, so linearity makes the
incentive constraint bind, as in
Lemma~\ref{lem:extreme-test-reduction}.

A common-rate test with rate \(\ell\) is incentive compatible exactly
when \(\ell(H-F)\le c\). Its payoff and its derivative in \(\ell\) are
\[
\frac{y(\beta)-k+\ell\beta G}{r+\ell},
\qquad
\frac{k-y_B+\beta(rG-\Delta_y)}{(r+\ell)^2}>0.
\]
The optimal common rate is therefore \(\lambda_S\), with the unchanged
bad-type value \(V_S\) and principal value
\eqref{eq:productive-stationary-payoff}. The latter crosses firing and
hiring at
\[
\beta_{S,0}^y=\frac{k-y_B}{\Delta_y+\lambda_S G},
\qquad
\beta_{S,h}^y=\frac{-B(r+\lambda_S)-k+y_B}
{r(G-B)-\lambda_SB-\Delta_y}.
\]
Both denominators are positive. Moreover, \(\Pi_S^y\) is increasing,
\(\Pi_S^y-h\) is decreasing, \(\Pi_S^y(0)<0\), and
\(\Pi_S^y(1)<G\). Since
\[
\Pi_S^y(\beta^*)=\frac{k_N^y(c)-k}{r+\lambda_S},
\]
the feasible stationary interval has nonempty interior exactly when
\(k<k_N^y(c)\), reduces to \(\{\beta^*\}\) at equality, and is empty
for \(k>k_N^y(c)\).

\medskip
\noindent
\textbf{Step 2: One-sidedness and stationary attachments.}
In posterior log-odds \(x\), the pass-only equations are
\[
V_x=-\frac{(H-V)rV}{c},
\qquad
\Pi_x=-\frac{(H-V)[r\Pi+k-y(\beta)]}{c}+\beta(G-\Pi),
\]
with \(\lambda_G=c/(H-V)\). The fail-only equations are
\[
V_x=\frac{(V-F)(rV+c)}{c},
\qquad
\Pi_x=\frac{(V-F)[r\Pi+k-y(\beta)]}{c}+(1-\beta)\Pi,
\]
with \(\lambda_B=c/(V-F)\). Set \(Q=\Pi-\Pi_S^y\), and use
\(\Phi_P\) and \(\Phi_F\) for the Hamiltonians of the two nonzero
extreme tests. Direct substitution gives
\[
\begin{aligned}
\Phi_P-\Phi_F
&=\frac{(H-F)(r+\lambda_S)}{V-F}Q
&&\text{in a pass-only region},\\
\Phi_F-\Phi_P
&=\frac{(H-F)(r+\lambda_S)}{H-V}Q
&&\text{in a fail-only region}.
\end{aligned}
\]
Define
\[
\Gamma_y(\beta,V)
:=\beta(rG+k-y_G)(V-F)-(1-\beta)(k-y_B)(H-V),
\]
and let \(\beta^{\circ,y}\in(0,1)\) be the unique solution of
\(\Gamma_y(\beta,V_S)=0\). Both partial derivatives of \(\Gamma_y\)
are positive on \(F<V<H\).

Binding incentives and admissibility make the rates continuous on each
nonstationary component. A mixed point therefore has a mixed neighborhood,
where \(Q=0\) and \(\Gamma_y=0\). Differentiating along this locus
and using the bad type's Bellman equation gives
\[
V_x=-\frac{(H-V)(V-F)}{H-F},
\qquad
\lambda_B-\lambda_G
=-\frac{[r(H-F)+c](V-V_S)}{2(H-V)(V-F)}.
\]
Mixed pass-learning points thus lie below \(\beta^{\circ,y}\), and
mixed fail-learning points lie above it.

Every nonstationary component has a stationary no-news endpoint. At pure
points the absolute log-odds drift exceeds \(\lambda_S\). On the mixed
locus it can vanish only at \((\beta^{\circ,y},V_S)\), which the preceding
side restrictions exclude as a no-news endpoint. Every interior no-news
endpoint is therefore reached in finite time. The proof of
Lemma~\ref{lem:learning-phases-wrong-stopping-action} carries over with
the modified flow. In particular, a firing endpoint of a pass-learning
component must lie at or below \(\beta_{S,0}^y\), and incentive
compatibility gives \(\limsup\lambda_G\le\lambda_S\). Below that
cutoff, \(y(\beta)-k+\beta\lambda_GG\) is strictly negative nearby.
At the cutoff, the same payoff-representation estimate gives
\(\Pi(\beta)=o(\beta-\beta_{S,0}^y)\), contradicting
\(\Pi\ge\Pi_S^y\) and \((\Pi_S^y)'>0\). A hiring endpoint of a
fail-learning component is excluded analogously, using
\[
\begin{aligned}
(\lambda_B-\lambda_G)(\Pi-h)_x
={}&[r+\beta\lambda_G+(1-\beta)\lambda_B](\Pi-h)\\
&+rh+k-y(\beta)+(1-\beta)\lambda_BB.
\end{aligned}
\]
The opposite stopping action makes the active pure rate diverge, while
the mixed locus excludes mixing nearby. At extreme beliefs, bounded
rates exclude mixed points. The only bounded-rate pure possibilities
are \(V\equiv0\) on a pass-learning left tail and \(V\to-c/r\),
when \(c+rF<0\), on a fail-learning right tail. The principal's equation
then gives, respectively,
\[
\Pi\longrightarrow\frac{y_B-k}{r}<0,
\qquad
\Pi\longrightarrow\frac{y_G-k}{r}<G,
\]
contradicting the lower envelope.

Consequently, \(k>k_N^y(c)\) rules out probation. At
\(k=k_N^y(c)\), no nonstationary region can attach to \(\beta^*\).
Mixing is excluded by stationary value matching and the mixed-point side
restrictions; pure attachments violate the lower envelope because
\[
\begin{aligned}
(\Pi-h)_x(\beta^*+)
&=-\frac{(H-V_S)[k-y(\beta^*)]}{c}<0,\\
\Pi_x(\beta^*-)
&=\frac{(V_S-F)[k-y(\beta^*)]}{c}>0,
\end{aligned}
\]
where \(k-y(\beta^*)=\lambda_S\beta^*G>0\). Only the optional
isolated stationary convention remains.

Suppose now that \(k<k_N^y(c)\). The lower envelope requires probation
at \(\beta^*\), so a stationary belief exists. Mixed points cannot
approach a stationary attachment \(s\): value matching would require
\(s=\beta^{\circ,y}\), contrary to the mixed-point side restrictions.
Each component is therefore pure near its attachment. The value gap obeys
\[
\begin{aligned}
Q_x&=\frac{\Gamma_y(\beta,V)}{r(H-F)+c}
-\left[\frac{r(H-V)}{c}+\beta\right]Q
&&\text{in a pass region},\\
Q_x&=\frac{\Gamma_y(\beta,V)}{r(H-F)+c}
+\left[\frac{r(V-F)}{c}+1-\beta\right]Q
&&\text{in a fail region}.
\end{aligned}
\]
Since \(Q(s)=0\) and \(Q\ge0\), a pass attachment requires
\(s\ge\beta^{\circ,y}\), and a fail attachment requires
\(s\le\beta^{\circ,y}\). The mixed-point restrictions then exclude
mixing throughout either component. The bad type's equations give
\(V_x>0\): \(V_S<V<0\) in a pass region, and
\(V>\max\{F,-c/r\}\) in a fail region. Hence \(\Gamma_y\) increases
strictly with \(x\), and the integrating-factor formula yields
\(Q>0\) away from the attachment, including when
\(s=\beta^{\circ,y}\). It also rules out a second zero at a finite
opposite endpoint.

There is a unique stationary belief. Two consecutive stationary beliefs
would force probation between them by the lower envelope, contradicting
the strict value gap on that component. A sole pass region must attach
at \(\beta_{S,0}^y\), a sole fail region at \(\beta_{S,h}^y\), and
two regions at \(\beta^{\circ,y}\). Two nondegenerate regions require
\(\beta_{S,0}^y<\beta^{\circ,y}<\beta_{S,h}^y\). Indeed, at
\(s=\beta_{S,0}^y\), we have \(k-y(s)=\lambda_S sG>0\), so a fail
branch would have \(\Pi_x(s^-)>0\) and violate the firing lower bound.
At \(s=\beta_{S,h}^y\), we have
\(rh(s)+k-y(s)=-\lambda_S(1-s)B>0\), so a pass branch would have
\((\Pi-h)_x(s^+)<0\) and violate the hiring lower bound. Equality
therefore omits the corresponding zero-length region.

\medskip
\noindent
\textbf{Step 3: Regime thresholds.}
Substitution of the stationary crossings into \(\Gamma_y\) gives
\[
\begin{aligned}
\beta_{S,0}^y\ge\beta^{\circ,y}
&\quad\Longleftrightarrow\quad k\ge k_P^y(c),\\
\beta_{S,h}^y\le\beta^{\circ,y}
&\quad\Longleftrightarrow\quad k\ge k_F^y.
\end{aligned}
\]
Moreover, the thresholds in \eqref{eq:productive-cost-thresholds} satisfy
\[
G[k_F^y-k_N^y(c)]=B[k_P^y(c)-k_N^y(c)].
\]
Since \(G>0>B\), the two regime thresholds lie on opposite sides of
\(k_N^y(c)\), unless all three coincide. Thus, within the active range
\(k<k_N^y(c)\), the pass-only and fail-only conditions cannot hold
simultaneously. If neither holds, both regions are present. Together with
stationary feasibility, these comparisons establish the stated regime
classification and its boundary conventions. All comparisons remain subject
to \eqref{eq:productive-maintained-restrictions} and the baseline condition
\(k>0\).

\medskip
\noindent
\textbf{Step 4: Construction and verification.}
The stationary values uniquely determine the solutions of the locally
Lipschitz systems in each region. Continuing a pass branch toward
\(\beta=1\) gives \(V\to0\) and
\[
\Pi\longrightarrow\frac{cG-H(k-y_G)}{c+Hr}<G.
\]
Its hiring gap is positive at the attachment, so a crossing exists. At
any crossing, \(Q>0\) implies \(h>\Pi_S^y\), and hence
\[
rh+k-y(\beta)>-\lambda_S(1-\beta)B>0,
\qquad
(\Pi-h)_x=-\frac{(H-V)[rh+k-y(\beta)]}{c}<0.
\]
The crossing is therefore unique and lies above \(\beta_{S,h}^y\).

For a fail branch, set \(Y=V-F\) and \(Z=\Pi/Y\). Then
\[
Y_x=\frac{Y(c+rF+rY)}{c},
\qquad
Z_x=\frac{k-y(\beta)}{c}-\left(\beta+\frac{rF}{c}\right)Z.
\]
As \(x\to-\infty\), if \(c+rF\ge0\), then \(Y\to0\), and
variation of constants gives \(Z\to(k-y_B)/(rF)<0\). If
\(c+rF<0\), then \(Y\to Y_\infty:=-(c+rF)/r>0\), and
\[
\Pi\longrightarrow
-\frac{(k-y_B)Y_\infty}{c+rY_\infty}<0.
\]
Since \(\Pi>0\) at the attachment, a firing crossing exists. At any
zero, \(Q>0\) implies \(\Pi_S^y<0\), so
\[
k-y(\beta)>\lambda_S\beta G>0,
\qquad
\Pi_x=\frac{Y[k-y(\beta)]}{c}>0.
\]
The crossing is unique and lies below \(\beta_{S,0}^y\).

Truncate each branch at its stopping crossing and stop outside the
resulting probation set. The crossings lie strictly inside \((0,1)\),
so rates are bounded. Both values match at the stationary no-news
attachment, and the stopping crossings make \(\Pi\) continuous. No
matching of \(V\) is required at the opposite stopping endpoints,
because no news moves away from them. The strict value gaps and crossing
locations verify the lower envelope globally. The payoff-representation
argument of Lemma~\ref{lem:bellman-characterization} identifies the
constructed solutions with the induced values,
using \(y(\beta)-k\) in place of \(-k\).

The Hamiltonian gaps select the prescribed nonzero extreme. It also
weakly dominates the feasible common test, whose Hamiltonian is
\(\lambda_S[\beta G-\Pi(\beta)]>0\), and hence dominates the zero
test. The bad type's unchanged Bellman equation and binding incentive
constraint allow Lemma~\ref{lem:global-effort-verification} to verify
shirking against all progressively measurable effort deviations. For
\(k\ge k_N^y(c)\), immediate stopping satisfies the lower envelope,
with the optional isolated stationary convention at equality. The
constructed assessments are therefore equilibria. The unique attachment,
local uniqueness of the region-specific equations, and transverse stopping
crossings establish uniqueness up to the baseline conventions.

Finally, if \(y_G=y_B=y\), the principal's flow payoff is
\(-(k-y)\), and every other payoff, signal law, and equilibrium condition
is unchanged. Since \(k-y>0\), the model is exactly the baseline with
this net probation cost.
\end{proof}

\begin{proof}[Proof of Proposition~\ref{prop:heterogeneous-phase-diagram}]
We adapt the proof of Theorem~\ref{thm:psio-phase-diagram}, retaining
its notation for the principal's and the bad type's values.

\medskip
\noindent
\textbf{Step 1: Local equations and stationary feasibility.}
Write \(b=\beta(1-\beta)(\lambda_B-\lambda_G)\). At nonstationary
probation beliefs, the Bellman equations and incentive constraint are
\[
\begin{aligned}
r_P\Pi&=-k+b\Pi'+\beta\lambda_G(G-\Pi)
                   -(1-\beta)\lambda_B\Pi,\\
r_AV&=bV'+\lambda_B(F-V),\\
\lambda_G(H-V)+\lambda_B(V-F)&\le c.
\end{aligned}
\]
Lemma~\ref{lem:extreme-test-reduction} applies unchanged, so every
optimal test exhausts the incentive constraint. A common-rate test is
feasible exactly when \(\ell(H-F)\le c\), independently of the
continuation value. Its payoff
\((\ell\beta G-k)/(r_P+\ell)\) increases with \(\ell\), giving the
binding rate \(\lambda_S\) and
\eqref{eq:heterogeneous-stationary-values}. The stationary payoff
crosses firing and hiring at
\[
\beta_{S,0}=\frac{k}{\lambda_S G},
\qquad
\beta_{S,h}=\frac{-B(r_P+\lambda_S)-k}
{r_P(G-B)-\lambda_SB}.
\]
The feasible stationary interval has nonempty interior exactly when
\(k<\lambda_S\beta^*G=k_N(c)\). At equality it reduces to
\(\{\beta^*\}\); for larger costs it is empty.

In posterior log-odds \(x\), a pass-only region uses
\(\lambda_G=c/(H-V)\) and satisfies
\[
V_x=-\frac{(H-V)r_AV}{c},
\qquad
\Pi_x=-\frac{(H-V)(r_P\Pi+k)}{c}+\beta(G-\Pi).
\]
A fail-only region uses \(\lambda_B=c/(V-F)\) and satisfies
\[
V_x=\frac{(V-F)(r_AV+c)}{c},
\qquad
\Pi_x=\frac{(V-F)(r_P\Pi+k)}{c}+(1-\beta)\Pi.
\]

\medskip
\noindent
\textbf{Step 2: Stationary attachments and one-sidedness.}
Set \(Q=\Pi-\Pi_S\) and define
\[
\Gamma_{A,P}(\beta,V)
:=\beta(r_PG+k)(V-F)-(1-\beta)k(H-V).
\]
Both partial derivatives are positive on \(F<V<H\). Its unique zero
at \(V=V_S\) is
\[
\beta^\circ=
\frac{k(Hr_A+c)}{k[r_A(H-F)+c]-FG\,r_Ar_P}.
\]
Using the nonzero extreme-test Hamiltonians \(\Phi_P\) and \(\Phi_F\)
from the baseline proof gives
\[
\begin{aligned}
\Phi_P-\Phi_F
&=\frac{(H-F)(r_P+\lambda_S)}{V-F}Q
&&\text{in a pass-only region},\\
\Phi_F-\Phi_P
&=\frac{(H-F)(r_P+\lambda_S)}{H-V}Q
&&\text{in a fail-only region}.
\end{aligned}
\]
At a mixed point, continuity of the rates gives a mixed neighborhood,
where \(Q=0\) and \(\Gamma_{A,P}=0\). Differentiating along this
locus and using the bad type's Bellman equation yields
\[
V_x=-\frac{(H-V)(V-F)}{H-F},
\qquad
\lambda_B-\lambda_G
=-\frac{[r_A(H-F)+c](V-V_S)}{2(H-V)(V-F)}.
\]
Thus mixed pass-learning points lie below \(\beta^\circ\), and mixed
fail-learning points lie above it.

Every nonstationary component has a stationary no-news endpoint.
To see this, at pure points the absolute log-odds drift exceeds
\(\lambda_S\). On the mixed locus it can vanish only at
\((\beta^\circ,V_S)\), which the preceding side restrictions exclude
as a no-news endpoint. Every interior no-news endpoint is therefore
reached in finite time. The argument of
Lemma~\ref{lem:learning-phases-wrong-stopping-action}, using the
respective discount rates, excludes firing at a pass-learning endpoint
and hiring at a fail-learning endpoint. The opposite stopping action
makes the active pure rate diverge, while the mixed locus excludes
mixing nearby. Nor can a no-news endpoint lie at an extreme belief.
The bounded-rate possibilities are \(V\equiv0\) on a pass-learning
left tail and \(V\to-c/r_A\), when \(c+r_AF<0\), on a fail-learning
right tail. In either case the principal's equation gives
\(\Pi\to-k/r_P\), contradicting the lower envelope. These are the
same endpoint exclusions as in the baseline proof.

Consequently, \(k>k_N(c)\) rules out probation. At \(k=k_N(c)\), no
nonstationary region can attach to \(\beta^*\). Mixing is excluded
by stationary value matching and the mixed-point side restrictions;
pure attachments violate the lower envelope because
\[
(\Pi-h)_x(\beta^*+)
=-\frac{(H-V_S)k}{c}<0,
\qquad
\Pi_x(\beta^*-)=\frac{(V_S-F)k}{c}>0.
\]
Thus only the optional isolated stationary convention remains.

Suppose now that \(0<k<k_N(c)\). The lower envelope requires probation
at \(\beta^*\), so a stationary belief exists. Mixed points cannot
approach a stationary attachment \(s\): value matching would require
\(s=\beta^\circ\), contrary to the mixed-point side restrictions.
Hence each component is pure near its attachment. Its value gap obeys
\[
\begin{aligned}
Q_x&=\frac{\Gamma_{A,P}}{r_P(H-F)+c}
-\left[\frac{r_P(H-V)}{c}+\beta\right]Q
&&\text{in a pass region},\\
Q_x&=\frac{\Gamma_{A,P}}{r_P(H-F)+c}
+\left[\frac{r_P(V-F)}{c}+1-\beta\right]Q
&&\text{in a fail region}.
\end{aligned}
\]
Since \(Q(s)=0\) and \(Q\ge0\), a pass attachment requires
\(s\ge\beta^\circ\), and a fail attachment requires
\(s\le\beta^\circ\). The mixed-point restrictions then exclude
mixing throughout either component. In a pass region,
\(V_S<V<0\); in a fail region, \(V>\max\{F,-c/r_A\}\). Hence
\(V_x>0\) and \(\Gamma_{A,P}\) increases strictly with \(x\) in
both regions. The integrating-factor formula gives \(Q>0\) away
from the attachment, including when \(s=\beta^\circ\), and rules
out a second stationary endpoint.

There is therefore a unique stationary belief: two consecutive
stationary beliefs would require probation between them by the lower
envelope, contradicting the strict value gap on that component.
A sole pass region attaches at \(\beta_{S,0}\), a sole fail region
at \(\beta_{S,h}\), and two regions attach at \(\beta^\circ\).
In the last case,
\(\beta_{S,0}<\beta^\circ<\beta_{S,h}\); equality omits the
corresponding zero-length region.

\medskip
\noindent
\textbf{Step 3: Regime thresholds.}
Let \(q\) be given by \eqref{eq:heterogeneous-composite-rate}, and define
\begin{equation}
k_P:=\frac{G(c+Fq)}{H-F},
\qquad
k_F:=\frac{BFq}{H-F}.
\label{eq:heterogeneous-region-thresholds}
\end{equation}
Substitution into \(\Gamma_{A,P}\) gives
\[
\beta_{S,0}\ge\beta^\circ
\quad\Longleftrightarrow\quad k\ge k_P,
\qquad
\beta_{S,h}\le\beta^\circ
\quad\Longleftrightarrow\quad k\ge k_F.
\]
Moreover,
\[
G[k_F-k_N(c)]=B[k_P-k_N(c)].
\]
Since \(G>0>B\), the region thresholds lie on opposite sides of
\(k_N(c)\), unless both equal it. Thus, for \(0<k<k_N(c)\), the
equilibrium is pass-only when \(k\ge k_P\), fail-only when
\(k\ge k_F\), and fail--pass when \(k<\min\{k_P,k_F\}\).
The first two conditions cannot hold simultaneously. Equality with
an active region threshold omits its zero-length region. These are
exactly the baseline regime conditions with \(r=q\).

For fixed discount rates,
\[
\frac{d}{dc}\left(\frac{q}{c}\right)
=-\frac{r_A[c^2+2c(H-F)r_P+(H-F)^2r_Ar_P]}
{c^2[c+(H-F)r_A]^2}<0.
\]
The ratio decreases from infinity to zero. Hence \(k_P\) crosses zero
exactly once, and the two regime thresholds cross \(k_N(c)\) together
exactly once, at a larger effort cost. Their ordering is therefore
the same as in the baseline phase diagram.

\medskip
\noindent
\textbf{Step 4: Construction and verification.}
The region-specific equations are locally Lipschitz on \(F<V<H\),
so the stationary values determine each branch uniquely. Continuing
a pass branch toward \(\beta=1\) gives \(V\to0\) and
\[
\Pi\longrightarrow\frac{cG-Hk}{c+Hr_P}<G.
\]
Its hiring gap is positive at the attachment, so a hiring crossing
exists. At every such crossing,
\[
(\Pi-h)_x=-\frac{(H-V)(r_Ph+k)}{c}<0,
\]
where \(h=\Pi>\Pi_S\ge0\). Hence the crossing is unique and lies
above \(\beta_{S,h}\).

For a fail branch, set \(Y=V-F\) and \(Z=\Pi/Y\). Then
\[
Y_x=\frac{Y(c+Fr_A+r_AY)}{c},
\qquad
Z_x=\frac{k}{c}
+\left[-\beta-\frac{Fr_A}{c}
       +\frac{(r_P-r_A)Y}{c}\right]Z.
\]
If \(c+Fr_A\ge0\), then \(Y\to0\) as \(x\to-\infty\), and
variation of constants gives \(Z\to k/(Fr_A)<0\). If
\(c+Fr_A<0\), then \(Y\to Y_\infty:=-(c+Fr_A)/r_A>0\), and
\[
\Pi\longrightarrow
-\frac{kY_\infty}{c+r_PY_\infty}<0.
\]
Since \(\Pi>0\) at the attachment, a firing crossing exists.
At every zero, \(\Pi_x=Yk/c>0\), so the crossing is unique and
lies below \(\beta_{S,0}\).

Truncate each branch at its stopping crossing and stop outside the
resulting probation set. The crossings lie strictly inside \((0,1)\),
so the rates are bounded. The values match at the stationary
attachment, and the stopping crossings make \(\Pi\) continuous.
The strict value gaps and crossing locations verify the lower
envelope globally, as in the baseline construction. The argument of
Lemma~\ref{lem:bellman-characterization}, using $r_P$ and $r_A$ for the
respective players, identifies the constructed functions with their induced
values. The Hamiltonian gaps select the prescribed nonzero extreme; its Hamiltonian is
\(r_P\Pi+k>0\), so it also dominates the zero test.
Lemma~\ref{lem:global-effort-verification}, with \(r_A\) in place
of \(r\), verifies shirking against all progressively measurable
effort deviations. For \(k\ge k_N(c)\), immediate stopping satisfies
the lower envelope, with the optional isolated stationary belief at
equality. These assessments are equilibria. The unique attachment,
local uniqueness of the branch equations, and transverse stopping
crossings establish uniqueness up to the baseline conventions.

\medskip
\noindent
\textbf{Step 5: Patience and the direction of monitoring.}
Fix \(c,k>0\) with \(k<k_N(c)\). Holding \(r_A>0\) fixed,
\(q\to\infty\) as \(r_P\to\infty\). Thus
\(k_P\to-\infty\) and \(k_F\to\infty\), giving pass-only
monitoring for every sufficiently large finite \(r_P\).

Holding \(r_P>0\) fixed, \(q\to0\) as \(r_A\downarrow0\).
The same limit holds along any sequence of positive discount rates
\((r_{A,n},r_{P,n})\to(0,0)\), since fixed \(c>0\) gives
\[
\frac{q(r_{A,n},r_{P,n};c)}{r_{A,n}}
=\frac{1+(H-F)r_{P,n}/c}{1+(H-F)r_{A,n}/c}
\longrightarrow1.
\]
In either case,
\[
k_F\longrightarrow0,
\qquad
k_P\longrightarrow\frac{Gc}{H-F}>k_N(c).
\]
Since \(0<k<k_N(c)\) is fixed, the fail-only regime conditions
hold eventually. The joint limit requires no restriction on the
relative convergence rates. For \(k\ge k_N(c)\), the absence of
nondegenerate probation, including the boundary convention at
equality, is unchanged along every sequence.
\end{proof}

\fi 

\end{document}